\documentclass[12pt,epsfig,a4]{article} 
\newcommand{\vs}{\vspace{-0.25cm}}
\usepackage{amsmath,graphicx}
\begin{document} 
\begin{center}
{\Large{\bf Radiative corrections to the charged pion-pair production process 
{\boldmath$\pi^-\gamma\to \pi^+\pi^-\pi^-$} at low energies}\footnote{This work 
has been supported in part by DFG and NSFC (CRC110).}  }  
\medskip

 N. Kaiser and S. Petschauer \\
\medskip
{\small Physik-Department T39, Technische Universit\"{a}t M\"{u}nchen,
   D-85747 Garching, Germany}
\end{center}
\medskip
\begin{abstract}
We calculate the one-photon loop radiative corrections to the charged pion-pair 
production process $\pi^-\gamma\to\pi^+\pi^-\pi^-$. In the low-energy region
this reaction is governed by the chiral pion-pion interaction. 
The pertinent set of 42 irreducible photon-loop diagrams is calculated by using 
the package FeynCalc. Electromagnetic counterterms with two independent 
low-energy constants $\widehat k_1$ and $\widehat k_2$ are included in order to 
remove the ultraviolet divergences generated by the photon-loops. Infrared 
finiteness of the virtual radiative corrections is achieved by including 
soft photon radiation below an energy cut-off $\lambda$. The purely 
electromagnetic interaction of the charged pions mediated by one-photon exchange 
is also taken into account. The radiative corrections to the total cross section 
(in the isospin limit) vary between $+10\%$ close to threshold and about $-1\%$ at 
a center-of-mass energy of $7m_\pi$. The largest contribution comes from the simple
one-photon exchange. Radiative corrections to the $\pi^+\pi^-$ and  
$\pi^-\pi^-$ mass spectra are studied as well. The Coulomb singularity of the 
final-state interaction produces a kink in the dipion mass spectra. The virtual 
radiative corrections to elastic $\pi^-\pi^-$ scattering are derived additionally. 
\end{abstract}



\section{Introduction and summary}
The pions ($\pi^+,\pi^0,\pi^-$) are the Goldstone bosons of spontaneous chiral 
symmetry breaking in QCD. Their strong interaction dynamics at low energies 
can therefore be calculated systematically (and accurately) with chiral 
perturbation theory in the form of a loop expansion based on an effective chiral
Lagrangian. The very accurate two-loop predictions \cite{cola} for the S-wave 
$\pi\pi$-scattering lengths, $a_0=(0.220\pm 0.005)m_\pi^{-1}$ and $a_2=(-0.044
\pm 0.001)m_\pi^{-1}$, have been confirmed experimentally by analyzing the 
$\pi\pi$ final-state interaction effects occurring in various (rare) charged 
kaon decay modes \cite{bnl,batley,cusp,batgamow}. Electromagnetic processes with pions 
offer further possibilities to test chiral perturbation theory. For example, 
pion Compton scattering $\pi^- \gamma \to\pi^- \gamma$ allows one to extract 
the electric and magnetic polarizabilities ($\alpha_\pi$ and $\beta_\pi$) of the 
charged pion. Chiral perturbation theory at two-loop order gives for the 
dominant pion polarizability difference the firm prediction $\alpha_\pi-\beta_\pi
=(5.7\pm1.0)\cdot 10^{-4}\,$fm$^3$ \cite{gasser}. It is however in conflict with 
the existing experimental results from Serpukhov $\alpha_\pi-\beta_\pi=(15.6\pm 
7.8)\cdot 10^{-4}\,$fm$^3$ \cite{serpukov1,serpukov2} and MAMI $\alpha_\pi-
\beta_\pi=(11.6\pm 3.4)\cdot 10^{-4}\,$fm$^3$ \cite{mainz}, which amount to 
values more than twice as large. 
In that contradictory situation it is promising that the COMPASS 
experiment \cite{private} at CERN aims at measuring the pion 
polarizabilities, $\alpha_\pi$ and $\beta_\pi$, with high statistics using the 
Primakoff effect. The scattering of high-energy negative pions in the 
Coulomb-field of a heavy nucleus (of charge $Z$) gives access to cross 
sections for $\pi^-\gamma$ reactions through the equivalent photon method 
\cite{pomer}. The theoretical framework to extract the pion 
polarizabilities from the measured cross sections for low-energy pion 
Compton scattering $\pi^- \gamma \to\pi^- \gamma$ or the primary pion-nucleus 
bremsstrahlung process $\pi^- Z \to\pi^- Z \gamma$ has been described (in the
one-loop approximation) in refs.\,\cite{picross,comptcor}. In addition to the 
strong interaction effects, the QED radiative corrections to real and virtual 
pion Compton scattering $\pi^-\gamma^{(*)} \to \pi^- \gamma$ have been 
calculated in refs.\,\cite{comptcor,bremscor}. The relative smallness of the 
pion-structure effects in low-energy pion Compton scattering \cite{picross} 
makes it necessary to include such higher order electromagnetic corrections. 

The COMPASS experiment is set up to detect simultaneously various 
(multi-particle) hadronic final-states which are produced in the Primakoff 
scattering of high-energy pions. The cross sections of the $\pi^-\gamma\to 3\pi$ 
reactions in the low-energy region offer new possibilities to test the strong 
interaction dynamics of the pions as predicted by chiral perturbation theory. 
In a recent analysis by the COMPASS collaboration \cite{private,compass} the 
total cross section for the process $\pi^-\gamma\to \pi^+\pi^- \pi^-$ has been 
extracted in the region from threshold up to a $3\pi$-invariant mass of about 
$5m_\pi$, and good agreement with the prediction of chiral perturbation 
theory has been found. The analysis of the $\pi^-\pi^0 \pi^0$-channel is ongoing 
\cite{private} and the corresponding experimental results are eagerly awaited. 
On the theoretical side the production amplitudes for 
$\pi^- \gamma \to \pi^-\pi^0 \pi^0$ and $\pi^- \gamma \to \pi^+\pi^-\pi^-$ have 
been calculated at one-loop order in chiral perturbation theory \cite{3pion}. 
It has been found that the next-to-leading order chiral corrections enhance sizeably 
(by a factor $1.5 -1.8$) the total cross section for neutral pion-pair production 
$\pi^-\gamma \to\pi^-\pi^0\pi^0$, but leave the one for charged pion-pair 
production $\pi^-\gamma \to\pi^+\pi^-\pi^-$ almost unchanged in comparison to the 
tree approximation. Let us note that these calculations would be implicitly contained 
in the work by Ecker and Unterdorfer \cite{ecker}, where the processes $\gamma^* 
\to 4\pi$ have been studied in chiral effective field theory. In particular, the 
chiral resonance theory used in that work offers the possibility to extend the 
description of the $\pi^-\gamma \to 3\pi$ reactions to higher energies.

The QED radiative corrections to neutral pion-pair production  $\pi^-\gamma \to 
\pi^- \pi^0\pi^0$ have been computed recently in ref.\,\cite{neutral}. This 
calculation was simplified by the fact that the virtual photon-loops could be 
represented by a multiplicative correction factor $R\sim \alpha/2\pi$ to the 
tree-amplitude and the number of contributing diagrams was limited to one dozen.  
The purpose of the present work is to extend the calculation of QED radiative 
corrections to the (more complex) charged pion-pair production process $\pi^-\gamma 
\to \pi^+ \pi^-\pi^-$. We use the package FeynCalc \cite{feyncalc} to evaluate the 
difficult set of 42 irreducible photon-loop diagrams. Other contributions of 
the same order in $\alpha$, such as the one-photon exchange and the soft-photon 
bremsstrahlung, can still be given in concise analytical formulas. As a result 
we find that the radiative corrections to the total cross section and the 
dipion mass spectra in the isospin limit are of the magnitude of a few percent. 
The largest contribution is provided by the simple one-photon exchange, which 
reaches up to $8\%$ close to threshold. It is however partly compensated by the 
leading isospin-breaking correction arising from the charged and neutral pion 
mass difference. The Coulomb singularity of the $\pi^\pm \pi^-$ final-state 
interaction causes a kink in the invariant mass  spectra.
   
Let us clarify that the present analysis is not a complete calculation of all
effects of order ${\cal O}(e^2p^2)$ in chiral perturbation theory, since only
photon-loops are considered but not the electromagnetic effects induced in
pion-loops via the charged and neutral pion mass difference. The latter (subleading)
isospin-breaking corrections are expected to be of similar size as the 
``genuine'' radiative corrections studied in this work.

\section{Charged pion-pair production: T-matrix and cross 
section}
We start out with recalling the kinematical and dynamical description 
\cite{3pion} of the charged pion-pair production process: $\pi^-(p_1)+
\gamma(k,\epsilon\,) \to\pi^+(p_2) +\pi^-(q_1)+\pi^-(q_2)$. It is advantageous
to choose for the transversal real photon the Coulomb-gauge in the 
center-of-mass frame, which entails the conditions $\epsilon \cdot p_1 
=\epsilon \cdot k= 0$. These subsidiary conditions imply that all diagrams 
for which the photon $\gamma(k,\epsilon\,)$ couples solely to the incoming 
negative pion $\pi^-(p_1)$ vanish identically. Under this valid specification 
the T-matrix has the following general form:
\begin{equation} T_{\rm cm} = {2e \over f_\pi^2} \Big[ \vec \epsilon \cdot \vec q_1
\,  A_1 + \vec \epsilon \cdot \vec q_2 \,  A_2 \Big] \,,\end{equation}
where $f_\pi= 92.4$\,MeV denotes the pion decay constant and $e$ is the 
elementary charge. In the above decomposition $A_1$ and $A_2$ are two 
(dimensionless) production amplitudes, which depend on five independent 
(dimensionless) Mandelstam variables $(s,s_1,s_2,t_1,t_2)$, defined as:
\begin{eqnarray} s\,m_\pi^2 = (p_1+k)^2  \,, &&s_1m_\pi^2=(p_2+q_1)^2\,,  
\quad t_1m_\pi^2=(q_1-k)^2\,, \nonumber \\ &&  s_2m_\pi^2=(p_2+q_2)^2
\,,  \quad t_2m_\pi^2=(q_2-k)^2\,. \end{eqnarray}
In this adapted notation $\sqrt{s}\,m_\pi$ is the total center-of-mass energy
of the process, with $m_\pi=139.57\,$MeV the charged pion mass. The introduced set 
of variables is particularly suitable for describing the permutation of the two 
identical negative pions in the final state, via the interchanges 
$(s_1\leftrightarrow s_2,\, t_1\leftrightarrow t_2)$. The second amplitude $A_2$ 
introduced in eq.\,(1) is determined by the crossing relation:
\begin{equation} A_2(s,s_1,s_2,t_1,t_2)= A_1(s,s_2,s_1,t_2,t_1)\,, \end{equation}
and therefore it is sufficient to specify only the first amplitude $A_1(s,s_1,s_2,t_1,t_2)$. 

At low energies the reaction $\pi^- \gamma \to \pi^+ \pi^-\pi^-$ is governed by 
the chiral pion-pion interaction at leading order \cite{3pion}. It is
advantageous to parameterize the special-unitary matrix-field $U$ in the chiral 
Lagrangian ${\cal L}_{\pi\pi}$ through an interpolating pion-field $\vec \pi$ in 
the form $U =\sqrt{1-\vec \pi^{\,2} /f_\pi^2} + i \vec \tau \cdot \vec \pi/f_\pi$. 
This has the consequence that no $\gamma 4\pi$ and $\gamma\gamma 4\pi$ 
contact-vertices exist at leading order. The tree amplitude of chiral 
perturbation theory reads \cite{3pion}:
\begin{equation}A_1^{(\rm tree)} = {2s-2-s_1-s_2+t_1+t_2\over 3-s-t_1-t_2} +{s-s_1
-s_2+t_2 \over t_1-1} \,, \end{equation}
where in each term the numerator stems from the (off-shell) 
$\pi\pi$-interaction in the isospin limit (proportional to $f_\pi^{-2}$) and the
 denominator from a pion-propagator. The respective tree diagrams are shown in 
Fig.\,1 of ref.\,\cite{3pion} and these coincide with the diagrams in Fig.\,2 of 
the present paper when deleting the external self-energy corrections. Note that 
the prefactor $2e/f_\pi^2$ in eq.\,(1) collects all occurring coupling constants.    

Applying the flux factor $[2m_\pi^2(s-1)]^{-1}$ and a symmetry factor $1/2$, 
the total cross section for the reaction $\pi^- \gamma \to \pi^+ \pi^-\pi^-$
is obtained by integrating the (polarization-averaged) squared transversal 
T-matrix over the three-pion phase space:
\begin{equation}\sigma_{\rm tot}(s)={\alpha\, m_\pi^2\over 32\pi^3 f_\pi^4 (s-1)} 
\int\limits_{z^2<1}\!\!\!\!\!\int\!d\omega_1 d\omega_2 \int_{-1}^1\!dx\int_0^\pi \! 
d\phi \, \big|\hat k \times (\vec q_1 A_1+\vec q_2 A_2)\big|^2 \,.\end{equation} 
Here, $\omega_1$ and $\omega_2$ are the center-of-mass energies of the outgoing 
negative pions divided by $m_\pi$. In terms of the directional cosines  
$x=\hat k \cdot \hat q_1,\,y= \hat k \cdot \hat q_2,\, z= \hat q_1 \cdot \hat q_2$ 
the squared cross products in eq.\,(5) take the form: 
\begin{equation} (\hat k \times \vec q_1)^2 = q_1^2(1-x^2)\,, \quad  (\hat k
\times \vec q_2)^2 = q_2^2(1-y^2)\,, \quad   (\hat k \times \vec q_1)\cdot 
(\hat k \times \vec q_2)= q_1 q_2(z-x y)\,, \end{equation}
with $q_{1,2} = \sqrt{\omega_{1,2}^2 -1}$ the momenta of the outgoing 
negative pions divided by $m_\pi$, and the subsidiary relations: 
\begin{equation} q_1q_2 \,z =\omega_1 \omega_2-\sqrt{s}(\omega_1 +\omega_2) 
+{s+1 \over 2}  \,, \quad\quad y = xz +\sqrt{(1-x^2)(1-z^2)} \cos\phi \,.
\end{equation}
The Mandelstam variables $s_1,\,s_2,\,t_1,\,t_2$ follow in the center-of-mass 
frame as:
\begin{eqnarray} && s_1 = s+1-2 \sqrt{s}\, \omega_2\,, \quad  s_2 =  
s+1-2 \sqrt{s} \,\omega_1\,,\nonumber \\ && t_1 = 1+{1-s\over 
\sqrt{s}} (\omega_1-q_1 x) \,,\quad t_2 = 1+{1-s\over \sqrt{s}} 
(\omega_2-q_2 y)\,. \end{eqnarray}
The (bounded) integration region in the $\omega_1\omega_2$ plane is determined 
by the inequality $z^2<1$. It is straightforward to solve at fixed $\omega_1$ 
the quadratic equation for the upper and lower limit $\omega_2^\pm$ of the 
$\omega_2$-integration:

\begin{equation} \omega_2^\pm = {1\over 2} \Bigg(\sqrt{s}-\omega_1\pm q_1 \,
\sqrt{s-2\sqrt{s} \omega_1-3 \over s-2\sqrt{s} \omega_1+1} \,\Bigg)\,, \qquad 
{\rm for} \quad 1< \omega_1 < {s-3 \over 2\sqrt{s}}\,.\end{equation} 
The right hand side of eq.\,(5) allows to calculate also the dipion mass spectra 
of the process  $\pi^- \gamma \to \pi^+ \pi^-\pi^-$ by omitting one 
integration over an energy variable. 
\section{Radiative corrections}
In this section we present the radiative corrections of relative order $\alpha 
=e^2/4\pi=1/137.036$ to the charged pion-pair production process  $\pi^- \gamma 
\to \pi^+ \pi^-\pi^-$. These include the purely electromagnetic interaction 
of the charged pions mediated by one-photon exchange and the one-photon loop 
corrections to the tree level diagrams of chiral perturbation theory. 
Electromagnetic counterterms \cite{knecht} will be needed in order to eliminate 
the ultraviolet divergences generated by the virtual photon-loops. Infrared 
finiteness of the radiative corrections is achieved in the standard way by 
including soft-photon bremsstrahlung below an energy cutoff.      
\subsection{One-photon exchange}
\begin{figure}
\begin{center}
\includegraphics[scale=0.8,clip]{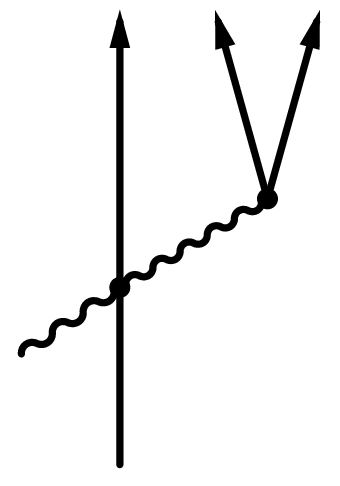}\quad\includegraphics[scale=0.8,clip]
{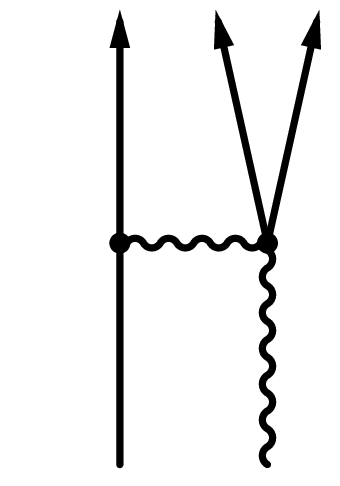}\\[1.0\baselineskip]\includegraphics[scale=0.8,clip]{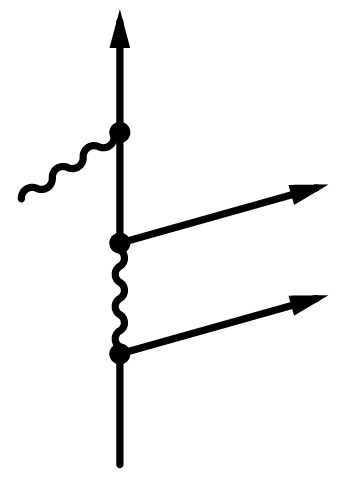}\quad
\includegraphics[scale=0.8,clip]{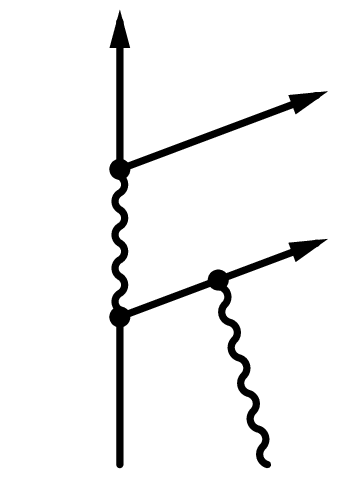}\quad\includegraphics[scale=0.8,clip]
{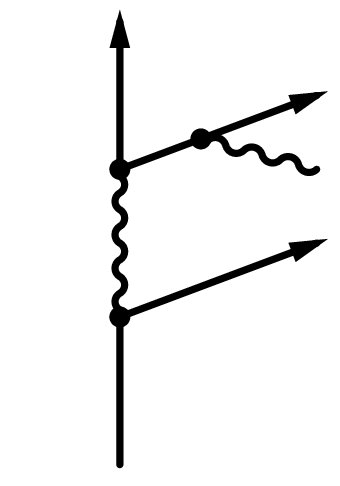}
\end{center}
\vspace{-.6cm}
\caption{Representative set of one-photon exchange diagrams for the process 
$\pi^-\gamma \to \pi^+ \pi^-\pi^-$. Each diagram has a partner due to the 
permutation of the two outgoing $\pi^-$. }
\end{figure}
The simplest electromagnetic correction to the process $\pi^- \gamma 
\to \pi^+ \pi^-\pi^-$ at low energies is given by describing the $\pi^-\pi^-$ 
interaction in terms of the one-photon exchange. A representative subset of the 10
corresponding tree diagrams is shown in Fig.\,1. The diagrams (in the upper 
row) involving the $\pi\pi\gamma\gamma$ contact-vertex of scalar quantum electrodynamics 
lead to the following production amplitude: 
\begin{equation}A_1^{(\gamma\gamma)} = 4\pi \alpha\, {f_\pi^2 \over m_\pi^2} \bigg\{ 
{2\over s_1}+ {1\over s_2}+ {1\over 2-s+s_2-t_1}\bigg\}\,, \end{equation}
while the additional diagrams with three ordinary $\pi\pi\gamma$ vertices (in the 
lower row) give rise to the production amplitude:
\begin{eqnarray}A_1^{(\gamma)} &=& 4\pi \alpha\, {f_\pi^2 \over m_\pi^2} \Bigg\{
{1\over 3-s-t_1-t_2}\bigg[{2s+1-s_1-2s_2+t_1 \over s-2-s_1+t_2}+{2s+1-2s_1-s_2
+t_2 \over s-2-s_2+t_1} \bigg] \nonumber \\ &&+{1\over t_1-1}
\bigg[ {1-2s+2s_1+s_2-t_1-2t_2 \over s_2}+ {s+1-s_1-2s_2+t_1+t_2 \over 
s-2-s_1+t_2}\bigg] \Bigg\}\,. \end{eqnarray}
The prefactor $f_\pi^2/m_\pi^2$ is a consequence of the 
normalization introduced in eq.\,(1).
\subsection{Photon-loop diagrams and electromagnetic counterterms}
\begin{figure}
\begin{center}
\includegraphics[scale=0.8,clip]{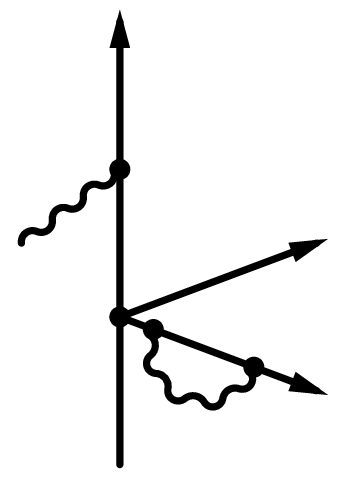}\quad\includegraphics[scale=0.8,clip]
{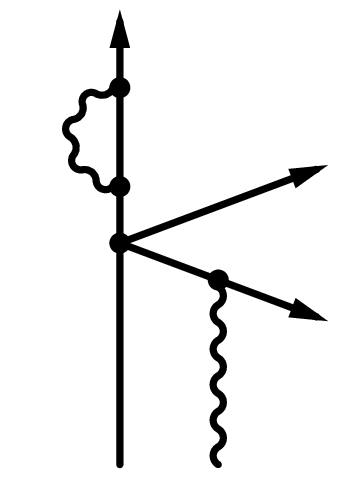}\quad\includegraphics[scale=0.8,clip]{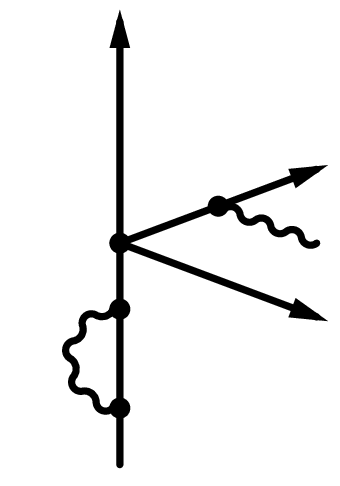}
\end{center}
\vspace{-.6cm}
\caption{Representative set of diagrams including wave function 
renormalization. For each diagram the self-energy correction can also be placed 
on the other three external legs.}
\end{figure}
The virtual radiative corrections to $\pi^-\gamma \to \pi^+ \pi^-\pi^-$
are obtained by dressing the (three) tree diagrams with a photon-loop in all 
possible ways. It is helpful to divide these loop diagrams into three classes: 
self-energy corrections on external pion-lines (I), vertex corrections to the 
pion-photon coupling (II), and ``irreducible'' photon-loop diagrams (III). 
We use dimensional 
regularization to treat both ultraviolet and infrared divergences (where the 
latter are caused by the masslessness of the photon). Divergent pieces of 
one-loop integrals show up in the form of the composite constant: 
\begin{equation} \xi = {1\over d-4} + {1\over 2}(\gamma_E-\ln 4\pi) +
\ln{m_\pi \over \mu_r} \,, \end{equation}
containing a simple pole at $d=4$. The arbitrary mass scale $\mu_r$ is introduced 
in order to keep (via a prefactor $\mu_r^{4-d}$) the mass dimension of one-loop 
integrals independent of $d$. Ultraviolet (UV) and infrared (IR) divergences are 
distinguished by the feature of whether $d<4$ or $d>4$ is the condition for
convergence of the $d$-dimensional integral. We discriminate them in the notation 
by putting appropriate subscripts, i.e. $\xi_{UV}$ and  $\xi_{IR}$. In order to
simplify all calculations we employ the Feynman gauge, where  the photon 
propagator is directly proportional to the Minkowski metric tensor $g_{\mu\nu}$.

Fig.\,2 shows a representative subset of the 12 photon-loop diagrams 
with a
self-energy correction on an external pion-line. The corresponding production 
amplitude  $A_1^{(\rm I)}$ is the tree amplitude in eq.\,(4) multiplied with twice the 
(electromagnetic) wave function renormalization factor $Z_2^{(\gamma)}-1$ of the 
charged pion: 
\begin{equation} A_1^{(\rm I)} = A_1^{(\rm tree)} \, {2 \alpha \over \pi }
(\xi_{IR} -\xi_{UV})\,. \end{equation}

\begin{figure}
\begin{center}
\includegraphics[scale=0.8,clip]{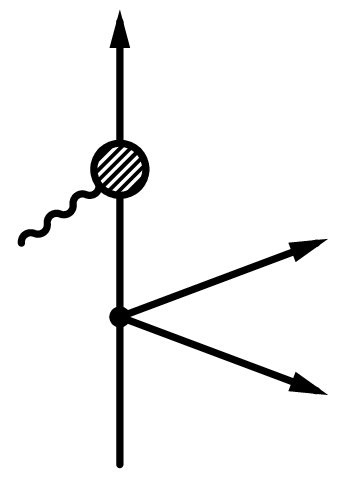}\quad\includegraphics[scale=0.8,clip]
{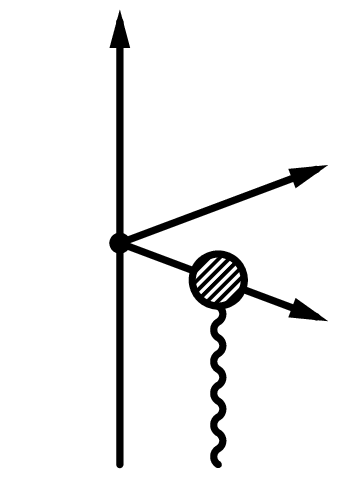}\quad\includegraphics[scale=0.8,clip]{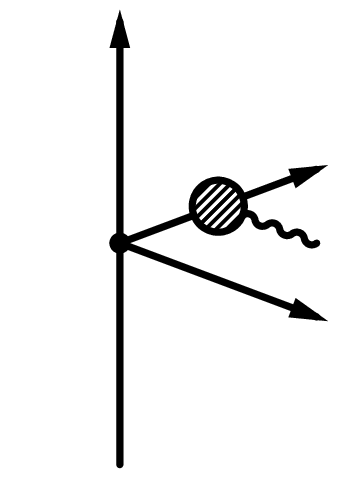}\\[1.0\baselineskip]
$\vcenter{\hbox{\includegraphics[scale=0.8,clip]{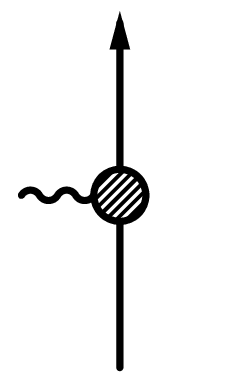}}}\ \Large{=}\ 
\vcenter{\hbox{\includegraphics[scale=0.8,clip]{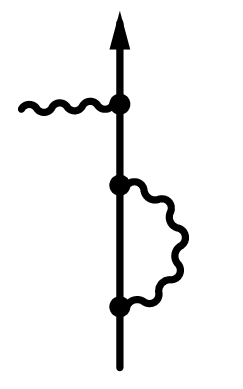}}}\ \Large{\bf +}\ 
\vcenter{\hbox{\includegraphics[scale=0.8,clip]{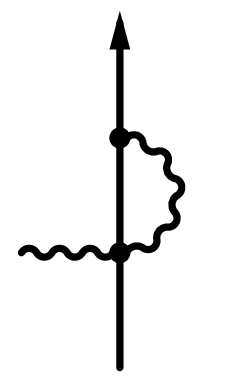}}} \Large{\bf +}\ 
\vcenter{\hbox{\includegraphics[scale=0.8,clip]{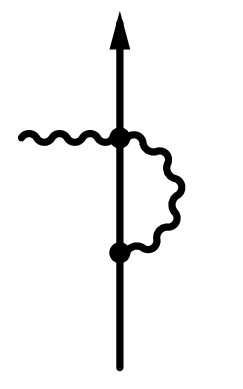}}} \ \Large{ +} \
\vcenter{\hbox{\includegraphics[scale=0.8,clip]{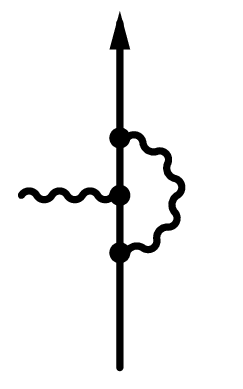}}}$
\end{center}
\vspace{-.6cm}
\caption{Diagrams including photonic vertex corrections. In the lower row the 
incoming pion is off-shell. }
\end{figure}

Fig.\,3 shows (reducible) photon-loop diagrams including vertex corrections 
to the pion-photon coupling. The blob introduced in the upper row stands for the 
sum of the four diagrams depicted in the lower row. This whole class of 
diagrams leads to a vanishing production amplitude:   
\begin{equation} A_1^{(\rm II)} =0\,, \end{equation}
where the zero results from the sum:
\begin{eqnarray}0 &=& {\alpha \over \pi} \bigg[-\xi_{UV}+1 -{u+1 \over 2u} 
\ln(1-u)\bigg] + {\alpha \over 8\pi}(6\xi_{UV}-7) \nonumber \\ && + {\alpha 
\over 8\pi} \bigg[6\xi_{UV}-6-{1\over u} +{u-1 \over u^2}(3u+1) \ln(1-u)\bigg] 
\nonumber \\ &&+ {\alpha \over 8\pi} \bigg[-4\xi_{UV}+5+{1\over u} +{u^2+6u+1 
\over u^2} \ln(1-u)\bigg] \,.  \end{eqnarray}
Here, each of the four summands corresponds to a diagram in the lower row of 
Fig.\,3 (in the order shown) and $u$ abbreviates the variable 
$(p_2-k)^2/m_\pi^2$,  $(q_1-k)^2/m_\pi^2$, or $(q_2-k)^2/m_\pi^2$ depending on which 
outgoing pion-line the vertex corrections take place. Note that the first term 
in eq.\,(15) is the once-subtracted (off-shell) self-energy of the pion. 
Nevertheless, it is most advantageous to combine it with the (proper) vertex 
corrections in Fig.\,3 in order to obtain the zero-sum. The same pattern of 
cancellation has been observed in section\,2 of ref.\,\cite{neutral}.  

\begin{figure}
\begin{center}
\includegraphics[scale=0.8,clip]{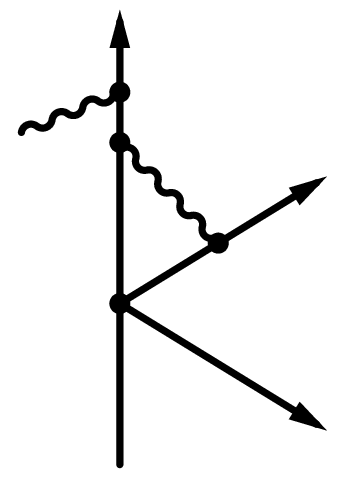}\quad\includegraphics[scale=0.8,clip]
{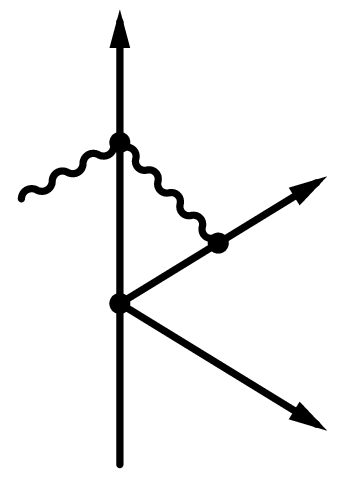}\quad\includegraphics[scale=0.8,clip]{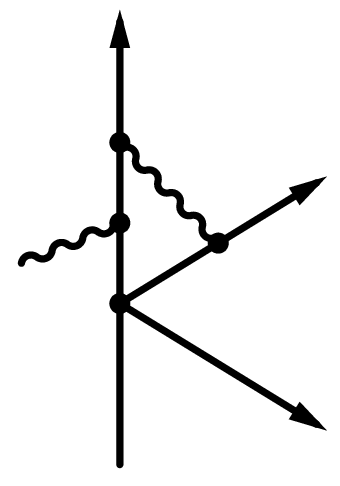}\\[1.0\baselineskip]
\includegraphics[scale=0.8,clip]{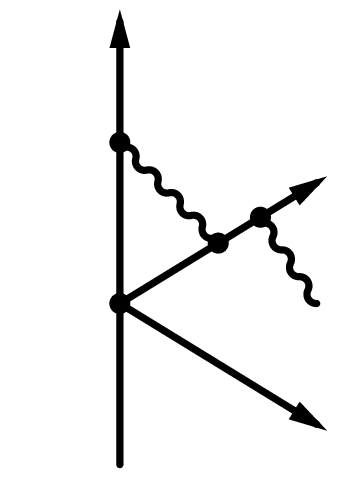}\quad\includegraphics[scale=0.8,clip]
{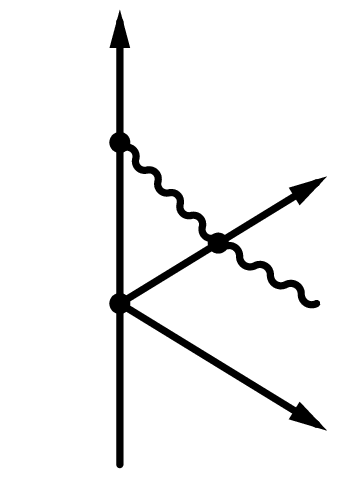}\quad\includegraphics[scale=0.8,clip]{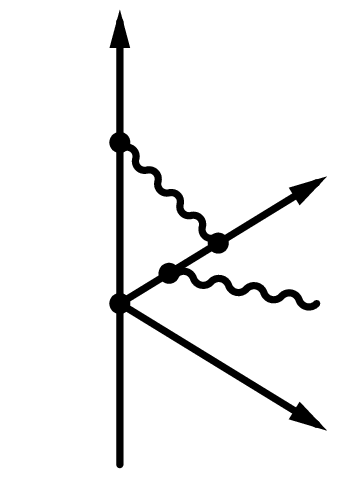}\quad\includegraphics[
scale=0.8,clip]{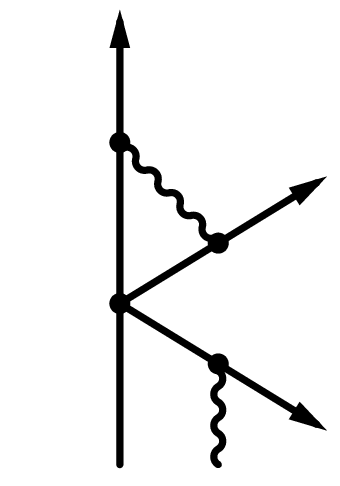}
\end{center}
\vspace{-.6cm}
\caption{Representative set of irreducible photon-loop diagrams for 
$\pi^-\gamma \to \pi^+ \pi^-\pi^-$.}
\end{figure}

In the remaining loop diagrams the virtual photon connects two charged 
pions with each other. Fig.\,4 shows for a selected pion-pair the 7 diagrams which 
result from all possible couplings of the incoming photon $\gamma(k,\epsilon\,)$. 
Given the 6 possible pion-pairs, there  are in total $6 \cdot 7 =42$ such 
``irreducible'' photon-loop diagrams for the process $\pi^-\gamma \to \pi^+ 
\pi^-\pi^-$. We have evaluated these diagrams by using the Mathematica package 
FeynCalc \cite{feyncalc}. After reduction to basic scalar loop-functions
the corresponding production amplitude $A_1$ consists of several hundreds of
terms, which prohibits a reproduction of the explicit analytical expression in 
this paper. A good check of the completeness of diagrams in the automatized 
calculation is provided by the crossing relation between $A_2$ and $A_1$ in 
eq.\,(3). The irreducible photon-loop diagrams generate also an ultraviolet 
divergent contribution. We have extracted this piece and after combining it with 
the $\xi_{UV}$-term from class\,I written in eq.\,(13) one gets in total the 
following ultraviolet divergent contribution from the virtual photon-loops:
\begin{equation}  A_1^{(\rm UV-div)} = {3 \alpha \over 2\pi}\, \xi_{UV}\bigg[ 
{2s-s_1-s_2+t_1+t_2 \over 3-s-t_1-t_2} +{s+2-s_1-s_2+t_2 \over t_1-1} \bigg]
 \,. \end{equation}
Note that $A_1^{(\rm UV-div)}$ is not proportional to the tree amplitude 
$A_1^{(\rm tree)}$ written in eq.\,(4).
Since chiral perturbation theory with inclusion of virtual photons is a
non-renormalizable effective field theory, the cancellation of ultraviolet 
divergences in radiative corrections requires the consideration of additional 
electromagnetic counterterms. The complete Lagrangian ${\cal L}_{e^2p^2}$ of order 
${\cal O}(e^2p^2)$ consists of 11 different terms and has been given in eq.\,(3.6) 
of ref.\,\cite{knecht}. We have extracted from ${\cal L}_{e^2p^2}$ all those vertices 
which are relevant for the process $\pi^-\gamma \to \pi^+ \pi^-\pi^-$ considered 
here. After evaluation of the tree diagrams shown in Fig.\,5 one obtains the 
following contribution to the production amplitude $A_1$ from the electromagnetic 
counterterms:
\begin{eqnarray} A_1^{(\rm ct)}&=&  {\alpha \over 2\pi}\bigg\{ (3 \xi_{UV}+ 
\widehat k_1 ) \bigg[{s_1+s_2-2s-t_1-t_2 \over 3-s-t_1-t_2}+{s_1+s_2-s-t_2-2 
\over t_1-1} \bigg] \nonumber \\ && \qquad - \,\widehat k_2 \,\bigg(
{1\over 3-s-t_1-t_2}+{1\over t_1-1} \bigg)\bigg\} \,,  \end{eqnarray}
with the two linear combinations of low-energy constants:
\begin{equation}\widehat k_1= {16 \pi^2 \over 9}(10 k_1^r -26k_2^r-54k_3^r
-27 k_4^r)-{3\over 2}-3\ln {m_\pi \over \mu_r}\,, \end{equation}
\begin{equation}\widehat k_2= {64 \pi^2 \over 9}(18 k_3^r +9k_4^r-5k_5^r
+31 k_6^r -k_7^r+36k_8^r)\,. \end{equation}
The last term $-3/2 -3 \ln(m_\pi/\mu_r)$ in eq.(18) comes from matching 
our convention for the ultraviolet divergence $\xi_{UV}$ to that of
ref.\,\cite{knecht}. One sees that in the sum $A_1^{(\rm UV-div)}+ A_1^{(\rm ct)}$ 
the ultraviolet divergence $\xi_{UV}$ drops out. This exact cancellation 
serves as a further important check of our calculation. It should be noted that 
the coefficients $\sigma_i$ written in eq.\,(3.11) of ref.\,\cite{knecht} (which 
determine the divergent part of an individual counterterm) are taken here 
consistently for $Z=0$, since we do not consider the additional electromagnetic 
effects induced in pion-loops via the charged and neutral pion mass difference.   
Actually, in the complete ChPT calculation to order ${\cal O}(e^2p^2)$ the 
low-energy constant $\widehat k_2$ in eq.\,(19) would be split into two pieces 
where one part is used to express the numerator of the chiral tree-amplitude 
in terms of the physical neutral pion mass square $m^2_{\pi^0}$ (see eq.\,(3.13) in 
ref.\,\cite{knecht}).    
\begin{figure}
\begin{center}
\includegraphics[scale=0.8,clip]{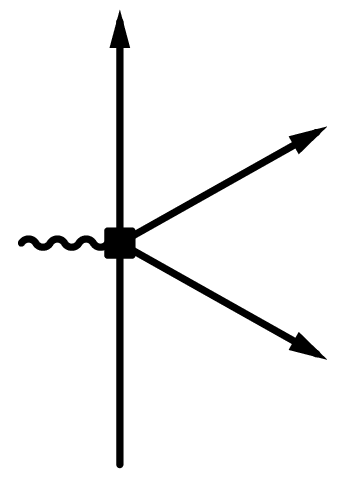}\quad\includegraphics[scale=0.8,clip]
{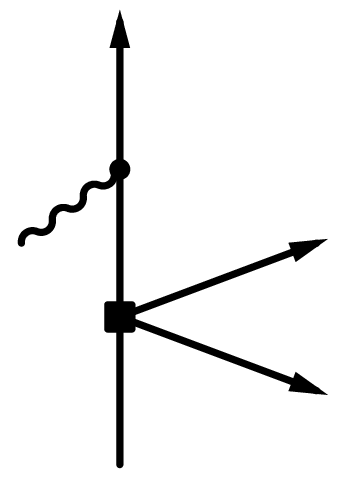}\quad\includegraphics[scale=0.8,clip]{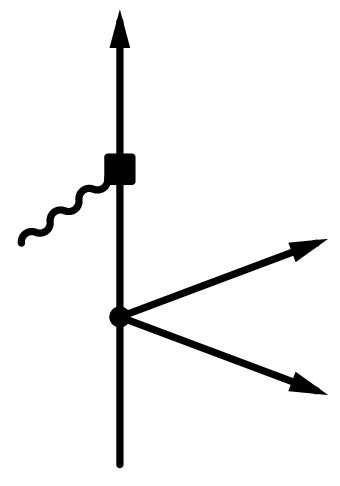}\quad\includegraphics[
scale=0.8,clip]{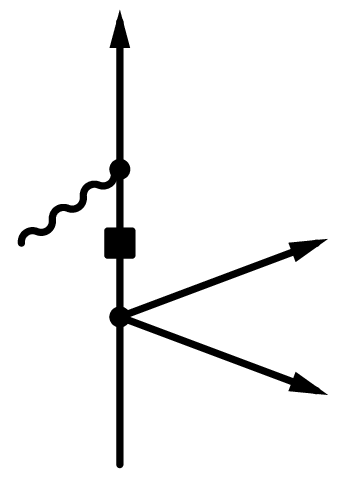}\\[1.0\baselineskip]\includegraphics[scale=0.8,clip]
{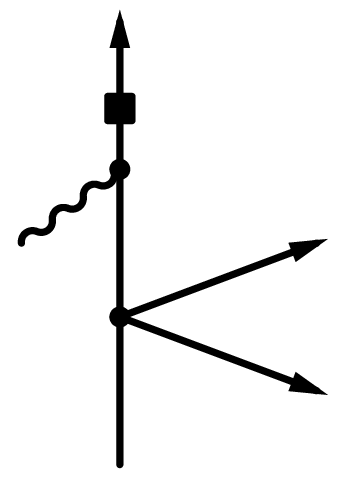}\quad\includegraphics[scale=0.8,clip]{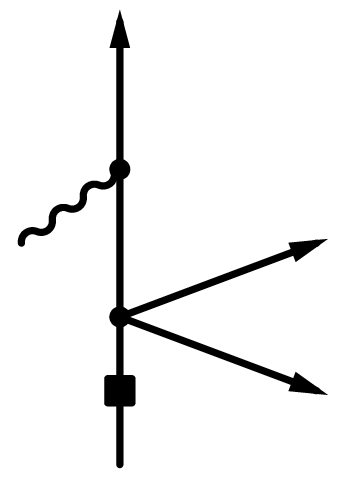}\quad\includegraphics[
scale=0.8,clip]{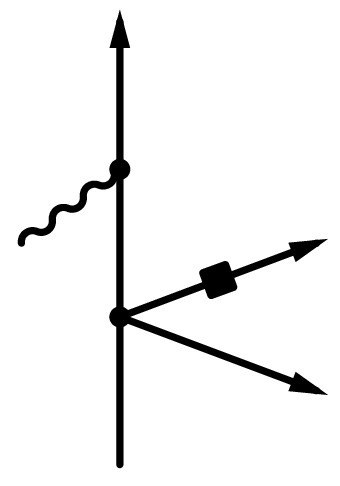}\quad\includegraphics[scale=0.8,clip]{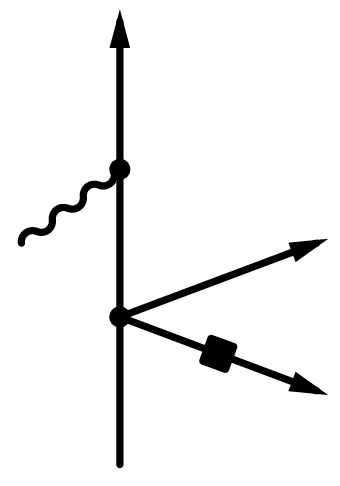}
\end{center}
\vspace{-.6cm}
\caption{Representative set of diagrams with electromagnetic counterterms 
symbolized by square-box vertices. The photon coupling to the other outgoing pions
leads to 14 additional diagrams. }
\end{figure}
\subsection{Soft photon bremsstrahlung}
In the next step we have to consider the infrared divergent terms proportional 
to $\xi_{IR}$ present in eq.\,(13) and in the production amplitude $A_1$ from the 
irreducible photon-loops. At the level of measurable cross sections these get 
eliminated by contributions from (undetected) soft-photon bremsstrahlung. In its 
final effect, soft-photon radiation off the in- or outgoing charged pions 
multiplies  the tree-level differential cross section for $\pi^-\gamma \to \pi^+ 
\pi^-\pi^-$ by the correction factor:
\begin{eqnarray} \delta_{\rm soft}&=& \alpha\, \mu^{4-d}\!\!\!\int\limits_{|\vec 
l\,|<\lambda}\!\!\!{d^{d-1}l  \over (2\pi)^{d-2}\, l_0} \bigg\{ {2p_1\cdot q_1 
\over p_1 \cdot l \, q_1 \cdot l} +{2p_2\cdot q_1 \over p_2\cdot l\,q_1 \cdot l} 
+{2p_1\cdot q_2 \over p_1 \cdot l \,q_2 \cdot l} +{2p_2\cdot q_2 \over p_2
\cdot l \, q_2 \cdot l}\nonumber \\ && - {2p_1\cdot p_2 \over p_1 \cdot l \, 
p_2 \cdot l} - {2q_1\cdot q_2 \over q_1 \cdot l \, q_2 \cdot l} - {m_\pi^2 \over 
(p_1 \cdot l)^2}- {m_\pi^2 \over (p_2 \cdot l)^2} - {m_\pi^2 \over (q_1 \cdot l)^2}
- {m_\pi^2 \over (q_2 \cdot l)^2}\bigg\} \,, \end{eqnarray}
which depends on a small energy cut-off $\lambda$. Working out this momentum 
space integral by the method of dimensional regularization (with $d>4$) one finds
two contributions. The first (universal) contribution includes the infrared 
divergence $\xi_{IR}$ and it has a logarithmic dependence on the cut-off 
$\lambda$. The detailed expression for $\delta_{\rm soft}^{(\rm uni)}$ reads:
\begin{eqnarray} \delta_{\rm soft}^{(\rm uni)}&=& {4\alpha\over \pi} \bigg( 
\ln{m_\pi \over 2\lambda} -\xi_{IR}\bigg)\bigg[(s_1+s_2+1-s-t_1-t_2)\,{\bf L}(
s_1+s_2-1-s-t_1-t_2) \nonumber \\ && +(2-s_1)\,{\bf L}(s_1-4) +(2-s_2)\,
{\bf L}(s_2-4)+  (s_2-s-t_1)\,{\bf L}(s+t_1-s_2-2) \nonumber \\ &&  
+(s_1-s-t_2)\,{\bf L}(s+t_2-s_1-2)+(s+1-s_1-s_2)\,{\bf L}(s-1-s_1-s_2) +1\bigg]
\,,\nonumber \\ \end{eqnarray}
with the logarithmic function:
\begin{equation} {\bf L}(z)={1\over \sqrt{z(4+z)}} \,\ln {\sqrt{z}+\sqrt{4+z}
\over 2}\,.  \end{equation}
Here, each of the six terms of the form $(\dots)\,{\bf L}(\dots)$ comes from an 
interference term in eq.\,(20) and the $1$ at the end from the sum of last four 
squares. The second contribution is specific for evaluating the soft-photon 
correction factor $\delta_{\rm soft}$ in the center-of-mass system and 
imposing an infrared cut-off, $|\vec l\,|<\lambda$, in this reference frame. 
Its explicit expression reads:
\begin{eqnarray} \delta_{\rm soft}^{(\rm cm)}&=&{\alpha\over 2\pi} \Bigg\{
{s+1 \over s-1} 
\ln s +\sum_{j=1}^3{2\omega_j \over \sqrt{\omega_j^2-1}}\,\ln\Big(\omega_j+\sqrt{
\omega_j^2-1}\,\Big) \nonumber \\ &&  +\sum_{j=1}^6 \int_0^1 dx \,{V_j C_j 
\over D_j \sqrt{C_j^2 -4s D_j}}\, \ln { C_j+ \sqrt{C_j^2 -4s D_j} \over  C_j- 
\sqrt{C_j^2 -4s D_j} }\Bigg\}\,, \end{eqnarray}
with 
\begin{equation}\omega_1 = {s+1-s_2 \over 2\sqrt{s}}\,,\qquad \omega_2 = 
{s+1-s_1 \over 2\sqrt{s}} \,,\qquad \omega_3= \sqrt{s}- \omega_1-\omega_2\,, 
\end{equation}
the center-of-mass energies of the outgoing pions divided by $m_\pi$, and the 
abbreviations:
\begin{equation*}V_1 = s+1-s_1-s_2\,, \qquad C_1 =  s+1-s_1+(s_1-s_2)x \,, 
\end{equation*}\vspace{-0.7cm}\begin{equation} D_1= 1+x(1-x)(s-1-s_1-s_2)\,,  \end{equation}
\begin{equation*} V_2 = s_1+s_2+1-s-t_1-t_2\,, \qquad C_2 =  s+1+(s_1+s_2-s-3)x \,,
 \end{equation*} 
\begin{equation} D_2= 1+x(1-x)(s_1+s_2-1-s-t_1-t_2)\,,\end{equation}
\begin{equation}V_3 = s_2-s-t_1\,, \qquad C_3 =  s+1 - s_2\, x \,,
 \qquad D_3= 1+x(1-x)(s+t_1-s_2-2)\,, \end{equation}
\begin{equation}V_4 = 2-s_1\,, \qquad C_4 =  s+1-s_2+(s_1+2s_2-s-3)x \,,
 \qquad D_4= 1+x(1-x)(s_1-4)\,, \end{equation}
\begin{equation}V_5 = s_1-s-t_2\,, \qquad C_5 =  s+1-s_1\, x \,,
 \qquad D_5= 1+x(1-x)(s+t_2-s_1-2)\,, \end{equation}
\begin{equation}V_6 = 2-s_2\,, \qquad C_6 =  s+1-s_1+(2s_1+s_2-s-3)x \,,
 \qquad D_6= 1+x(1-x)(s_2-4)\,, \end{equation}
for linear polynomials in the dimensionless Mandelstam variables 
$(s,s_1,s_2,t_1,t_2)$.
\section{Results and discussion}
In this section we present and discuss the numerical results for the radiative 
corrections to the charged pion-pair production process $\pi^-\gamma \to \pi^+ 
\pi^-\pi^-$. We study in detail the radiative corrections to the total cross 
section $\sigma_{\rm tot}(s)$ and to the $\pi^+\pi^-$ and  $\pi^-\pi^-$ 
invariant mass spectra. At the level of the production amplitudes the radiative 
corrections are given in eq.\,(5) by the real parts of interference terms between 
$A_{1,2}^{\rm (tree)}$ and $A_{1,2}^{\rm (rad-cor)} \sim \alpha$. The   
lengthy expressions generated by FeynCalc have been evaluated numerically 
with the help of the routine LoopTools \cite{looptools}. By assigning different 
values to the parameter $\xi_{IR}$ in the code the exact cancellation of infrared 
divergences from virtual photon-loops and soft-photon radiation (see eq.\,(21)) has 
been verified.    
\subsection{Radiative corrections to total cross section}
\begin{figure}[ht]
\begin{center}
\includegraphics[scale=0.5,clip]{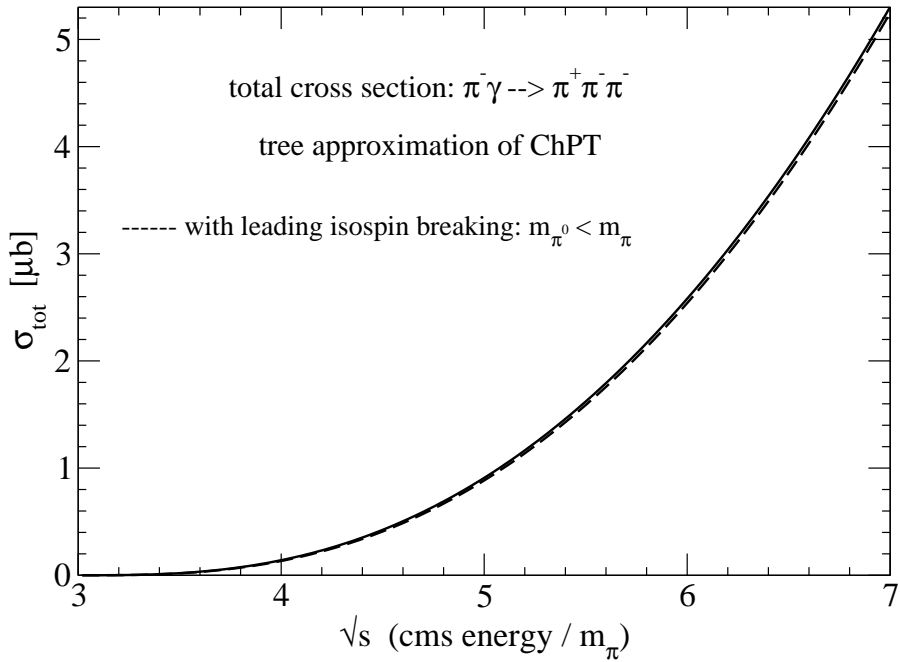}
\end{center}
\vspace{-.6cm}
\caption{Total cross section $\sigma_{\rm tot}(s)$ of the process $\pi^-\gamma \to 
\pi^+ \pi^-\pi^-$ calculated at tree-level in chiral perturbation theory. The full line 
corresponds to the isospin limit and the dashed line includes the leading isospin-breaking 
correction due to $m_{\pi^0}<m_\pi$.}
\end{figure}
For orientation, we reproduce in Fig.\,6 the total cross section 
$\sigma_{\rm tot}(s)$ of the reaction $\pi^-\gamma \to \pi^+ \pi^-\pi^-$ calculated  
at tree-level in chiral perturbation theory \cite{picross,3pion}. As demonstrated 
in ref.\,\cite{3pion} $\sigma_{\rm tot}(s)$ remains almost unchanged in the region 
$3<\sqrt{s}<6$ after inclusion of the next-to-leading order chiral corrections 
(from pion-loops and chiral counterterms in the isospin limit). The dashed line in 
Fig.\,6 is obtained by taking into account the leading isospin-breaking correction 
due to the difference of the charged and neutral pion mass (see eq.\,(2.6) in 
ref.\,\cite{knecht}). Isospin-breaking modifies the tree amplitude $A_1^{(\rm tree)}$ in 
such a way that the constant $\delta_{\rm ib} = 2(m_{\pi^0}/m_\pi)^2 -2 = -0.1295$ is 
added to both numerators in eq.\,(4). The relative correction to the total cross 
section  $\sigma_{\rm tot}(s)$ arising from this effect amounts 
to $-7.5\%,\, -5.0\%,\, -2.6\%,\, -1.6\%,\, -1.1\%$ at $\sqrt{s} =3.5,\,
4,\,5,\,6,\,7$, respectively (see Fig.\,8).   

Fig.\,7 shows in percent the radiative corrections to the 
total cross section $\sigma_{\rm tot}(s)$ in the region $3<\sqrt{s}<7$. The (lower) 
dashed-dotted and dashed curves display the effects of the soft-photon 
bremsstrahlung, separated into the universal contribution proportional to 
$\ln(m_\pi/2\lambda)$ and the contribution specific for imposing the infrared 
cut-off in the center-of-mass frame. As in refs.\,\cite{comptcor,neutral} the value 
$\lambda =5\,$MeV has been chosen, which is in the order of magnitude as it 
appears in the kinematics measured at the COMPASS experiment \cite{janhabil}.
The (upper) full line in Fig.\,7 shows the effect of the virtual photon-loops. 
Surprisingly, this radiative correction which requires a huge effort for its
computation is almost constant in the region $3.5<\sqrt{s}<7$. With an average 
value of $1.6\%$ it is also rather close to the analogous result (about $2.2\%$) 
for neutral pion-pair production $\pi^-\gamma \to \pi^- \pi^0\pi^0$ shown in 
Fig.\,5 of ref.\,\cite{neutral}. The behavior close to threshold  $3<\sqrt{s}<3.5$ 
is governed by the Coulomb singularity in the $\pi^\pm\pi^-$ final-state 
interaction, i.e. the Gamow factor $\pm \pi \alpha/\beta$ with $\beta$ the relative 
pion velocity \cite{batgamow}. 
Since the $\pi^+ \pi^-\pi^-$ final-state gives rise to two pion-pairs with opposite 
charges and only one with equal charges the attractive effect (i.e. an 
enhancement of the cross section) prevails. The effects of the Coulomb singularity
will be analyzed in more detail in subsection\,4.2 and in the appendix. 

Fig.\,8 shows by the (upper) dashed line the electromagnetic corrections arising 
from the one-photon exchange and the sum of the previous contributions 
(in Fig.\,7) is reproduced by the (lower) full curve. With values ranging 
between $8\pi \alpha f_\pi^2/m_\pi^2 = 8.04\%$ at threshold and $2.1\%$ at 
$\sqrt{s}=7$ the simple one-photon exchange constitutes the largest correction of 
relative order $\alpha$. The horizontal grey band shows the additional effects of 
the electromagnetic counterterms. We have varied the low-energy constants 
$ \widehat k_1$ and $ \widehat k_2$ in the range $ -1\leq \widehat k_{1,2} \leq 1$, 
which is expected to cover electromagnetic counterterms of natural size 
\cite{knecht}. Evidently, if one allows for a larger range of the low-energy 
constants $\widehat k_{1,2}$ the band will broaden accordingly. The total sum of the 
radiative corrections is shown by the dropping band in Fig.\,8. The dashed-dotted 
line is finally obtained by complementing this sum (for $\widehat k_{1,2}=0$) by 
the leading isospin-breaking correction.   
Putting aside the one-photon exchange contribution which is special for the 
process $\pi^-\gamma \to \pi^+ \pi^-\pi^-$, one can conclude that the 
radiative corrections calculated here are of similar size as in the case of the 
neutral pion-pair production process $\pi^-\gamma \to \pi^- \pi^0\pi^0$ studied in 
ref.\,\cite{neutral}. Moreover, one finds that the one-photon exchange and the 
leading isospin-breaking correction given by $\delta_{\rm ib} = 2(m_{\pi^0}/m_\pi)^2 
-2$ compensate each other partly.
\begin{figure}[t!]
\begin{center}
\includegraphics[scale=1.0,clip]{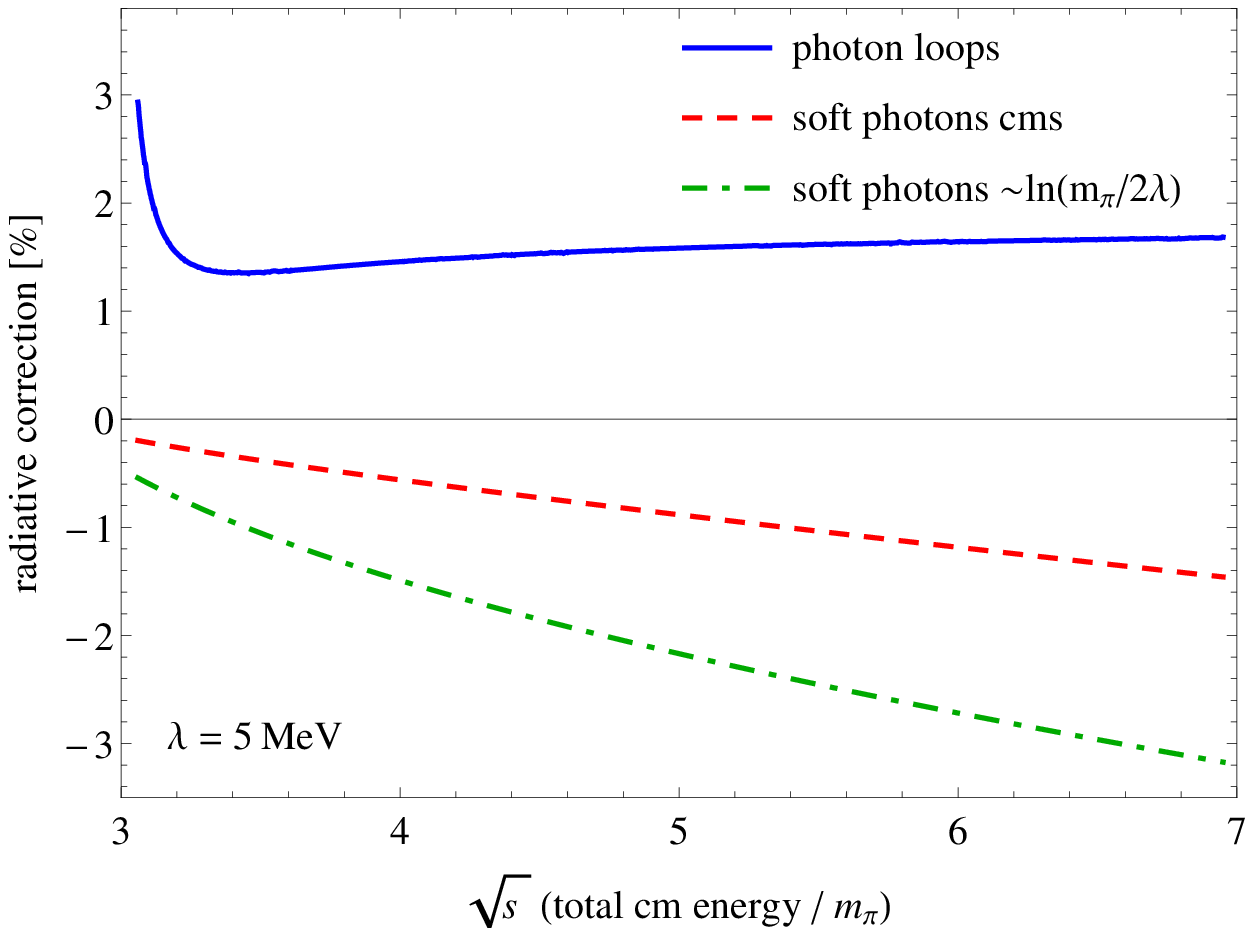}
\end{center}
\vspace{-.6cm}
\caption{Radiative corrections to the total cross section of the process 
$\pi^-\gamma \to \pi^+ \pi^-\pi^-$.}
\end{figure}
\begin{figure}[t!]
\begin{center}
\includegraphics[scale=1.0,clip]{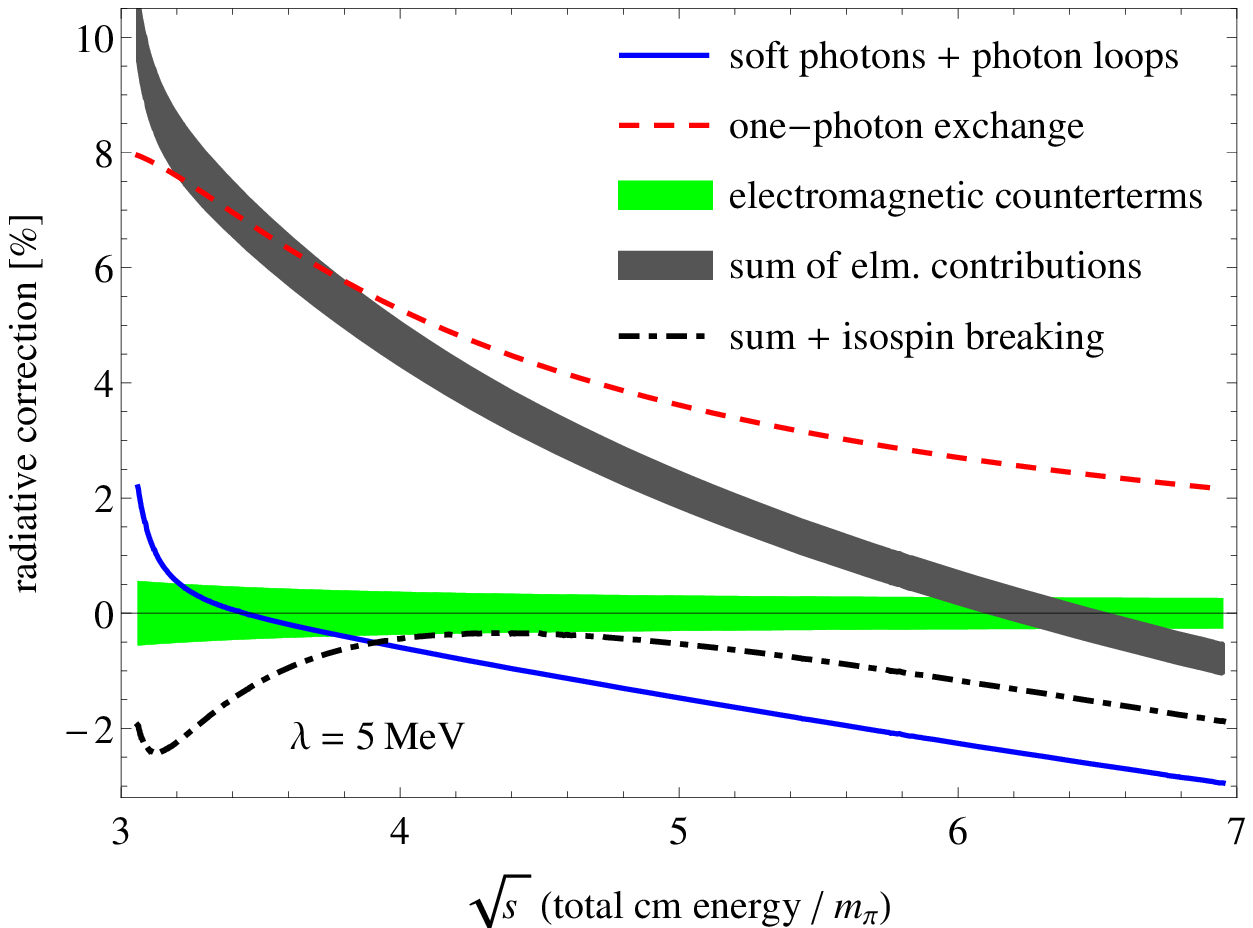}
\end{center}
\vspace{-.6cm}
\caption{Radiative corrections to the total cross section of the process 
$\pi^-\gamma \to \pi^+ \pi^-\pi^-$.}
\end{figure}

\subsection{Radiative corrections to dipion mass spectra}
\begin{figure}[t!]
\begin{center}
\includegraphics[scale=0.5,clip]{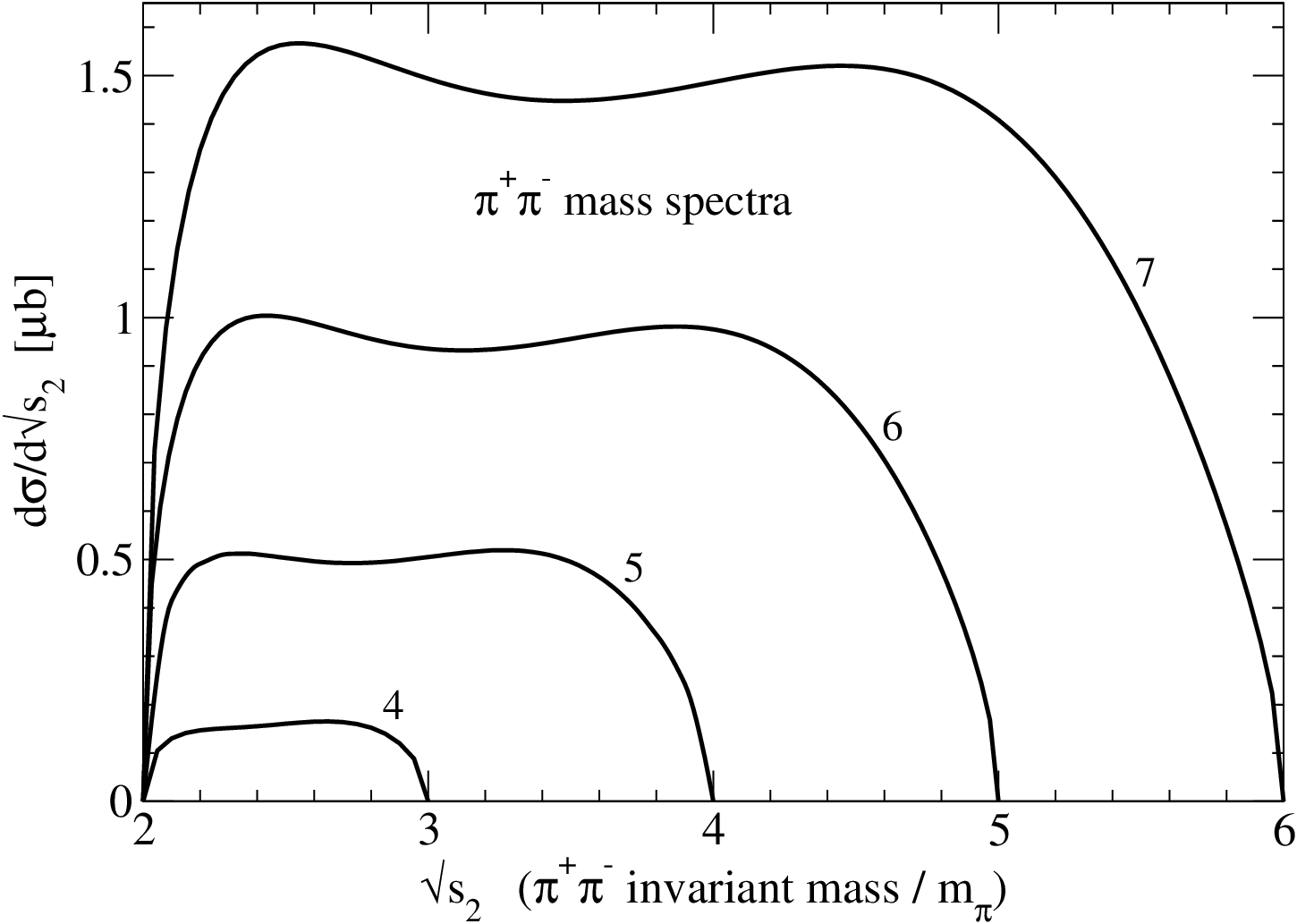}
\end{center}
\vspace{-.5cm}
\caption{$\pi^+\pi^-$ mass spectra of the process $\pi^-\gamma \to \pi^+ 
\pi^-\pi^-$ as a function of the $\pi^+\pi^-$ invariant mass $\sqrt{s_2} 
\,m_\pi $ for values of $\sqrt{s}=4,5,6,7$.} 
\end{figure}

In this subsection we study the radiative corrections to more exclusive 
observables of the process $\pi^-\gamma \to \pi^+ \pi^-\pi^-$, namely the 
dipion mass spectra. We start with the $\pi^+\pi^-$ mass spectrum. In terms of the
Mandelstam variables the $\pi^+\pi^-$ invariant mass is $\sqrt{s_2}\,m_\pi $. 
Exploiting the relation $s_2= s+1-2 \sqrt{s}\omega_1$, one sees that the 
differential cross section $d\sigma/d\sqrt{s_2}$ is obtained from the right 
hand side of eq.\,(5) by leaving out the integration over $\omega_1$ and multiplying 
with a factor $\sqrt{s_2/s}$.  Again for orientation, we reproduce in Fig.\,9 
the $\pi^+\pi^-$ mass spectrum calculated at tree-level in the isospin limit \cite{3pion}. The 
numbers next to the curves are the values of $\sqrt{s}$. The radiative corrections 
arising from soft-photon bremsstrahlung and virtual photon-loops are shown by the
dashed and solid curves in Fig.\,10. Note that these corrections are given 
here in absolute units of microbarn, without dividing by the respective mass spectra at 
tree-level. The pattern of radiative corrections is completed in Fig.\,11, where 
the effect the one-photon exchange and the sum of all contributions are shown. 
Throughout, one observes positive corrections to  $d\sigma/d\sqrt{s_2}$ from the 
virtual photon-loops and the one-photon exchange and negative corrections from 
the soft-photon radiation. In the total sum of all contributions oscillations 
and sign-changes occur, which get more pronounced with increasing $\sqrt{s}$.    
The overlying bands produced by the variation of the low-energy constants  
$\widehat k_1$ and $\widehat k_2$ have not been included for reasons of a 
cleaner presentation.     

\begin{figure}[t!]
\begin{center}
\includegraphics[scale=1.0,clip]{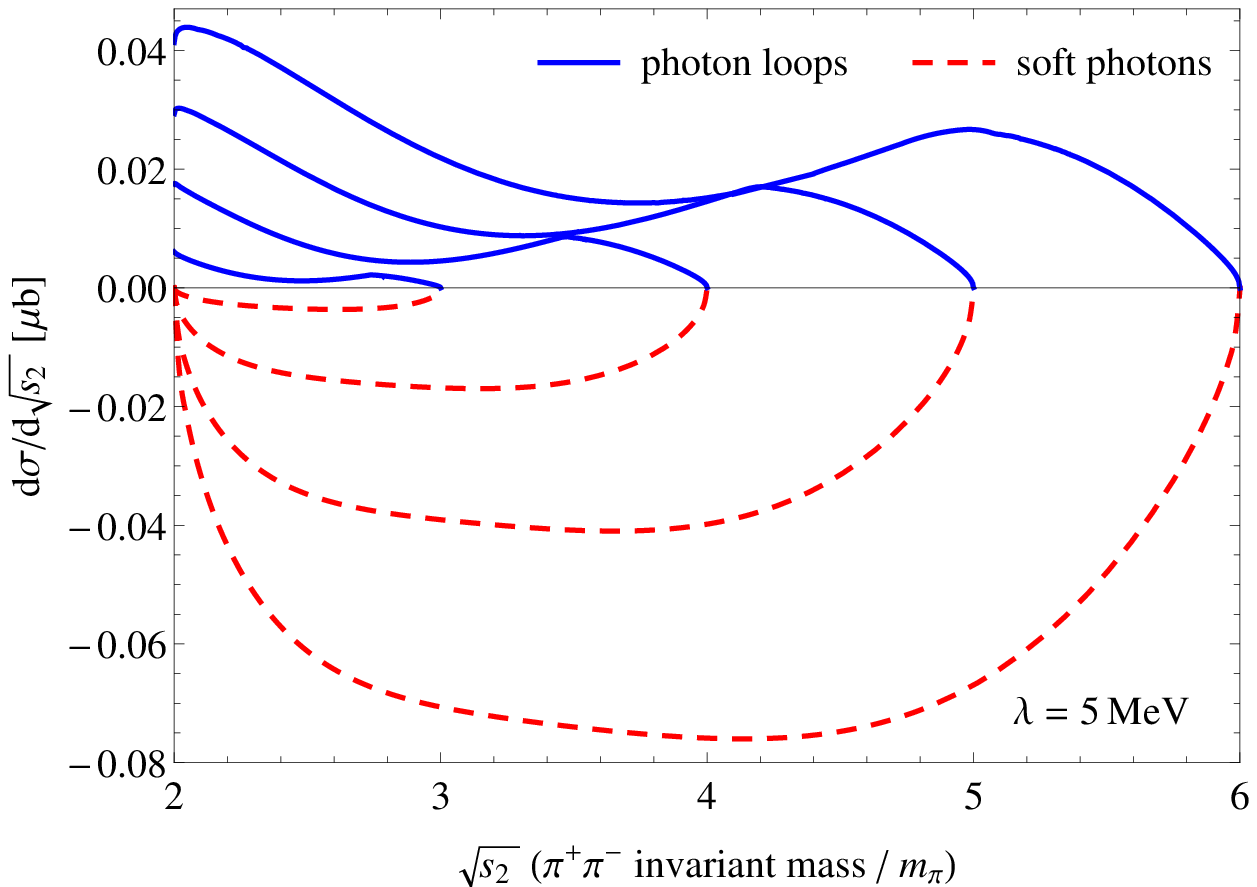}
\end{center}
\vspace{-.6cm}
\caption{Radiative corrections to the $\pi^+\pi^-$ mass spectrum of the process 
$\pi^-\gamma \to \pi^+ \pi^-\pi^-$ for $\sqrt{s}=4,5,6,7$.}
\end{figure}
\begin{figure}[t!]
\begin{center}
\includegraphics[scale=1.0,clip]{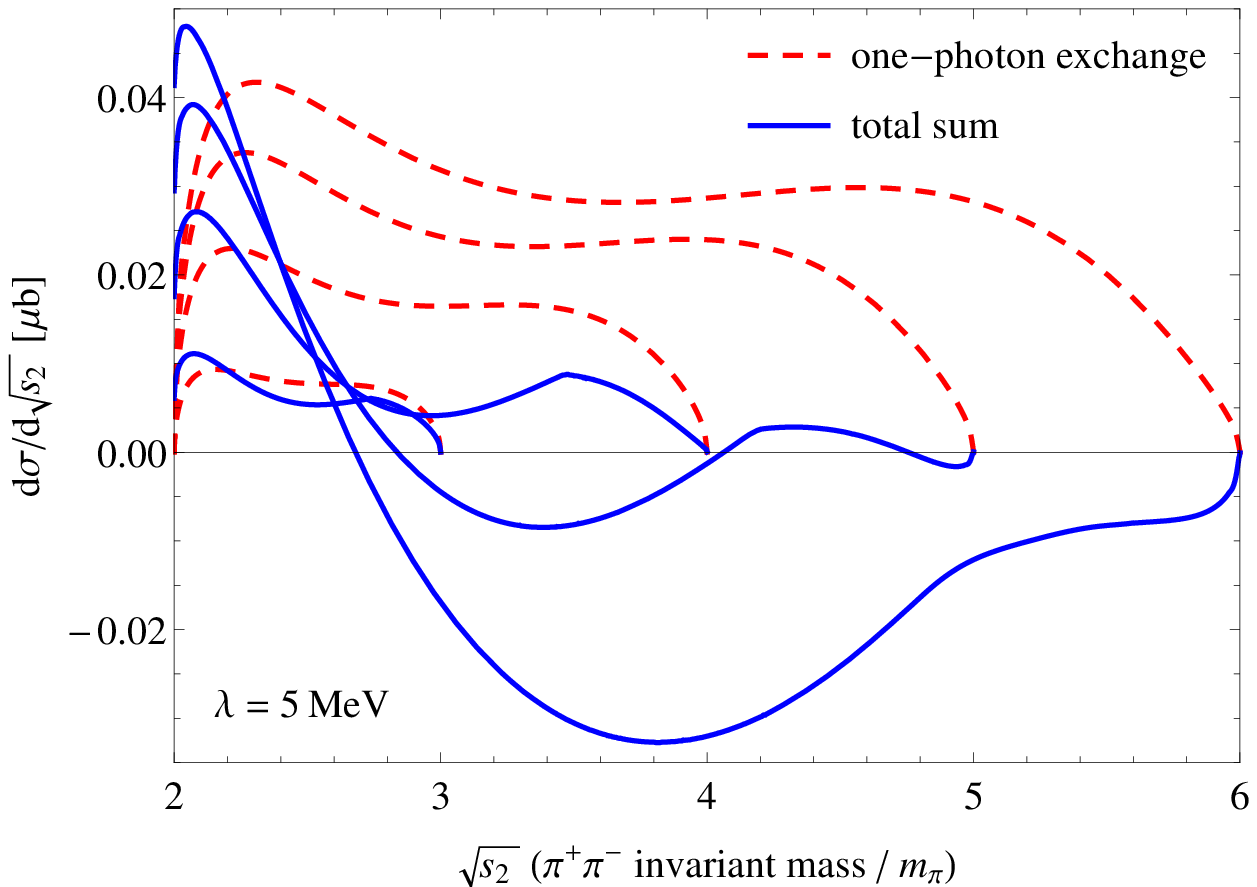}
\end{center}
\vspace{-.6cm}
\caption{Radiative corrections to the $\pi^+\pi^-$ mass spectrum of the process 
$\pi^-\gamma \to \pi^+ \pi^-\pi^-$ for $\sqrt{s}=4,5,6,7$.}
\end{figure}

A striking feature visible in Figs.\,10,\,11 is that the full curves are not smooth, 
but have kinks at intermediate values of $\sqrt{s_2}$. These kinks are by no means 
numerical artifacts, but they have a clear physical origin in the Coulomb 
singularity of the electromagnetic $\pi^\pm\pi^-$ final-state interaction. Let us 
elaborate on this close relationship in more detail.  At fixed $s_2$ the range of 
the (integration) variable $s_1$ is $s_1^- < s_1<s_1^+$ with:
\begin{equation} s_1^\pm =  {1\over 2} \Bigg( 3+s-s_2 \pm  \sqrt{s_2-4\over s_2}
\,\sqrt{(s+1-s_2)^2-4s}\,\Bigg)\,, \end{equation}
where $4<s_2<(\sqrt{s}-1)^2$. The isolated Coulomb singularity $1/\sqrt{s_1-4}$ leads to a dipion 
mass spectrum of the form:
\begin{equation} F(\sqrt{s_2}\,)= \int_{s_1^-}^{s_1^+}ds_1\, {1\over 
\sqrt{s_1-4}} = 2\sqrt{s_1^+ -4}-2\sqrt{s_1^- -4}\,. 
\end{equation}
The just constructed function $F(\sqrt{s_2}\,)$ is shown by the full lines 
in Fig.\,12 for $\sqrt{s}=4,5,6,7$. One observes a sharp kink at the position 
$s_2=(s-1)/2$, which is 
determined by the solution of the equation $s_1^- =4$. At this point the integral 
in eq.\,(32) extends fully into the inverse square-root singularity. One finds by 
a simple calculation that the left and right derivative of $F(\sqrt{s_2}\,)$ with 
respect to the variable $\sqrt{s_2}$ at this point are different with values: 
$\sqrt{2/(s-9)}\big[\pm 8-\sqrt{2(s-1)}\,\big]$. Therefore, this example 
demonstrates that the Coulomb singularity causes inevitably a kink in the two-pion 
mass spectrum. Actually, it is evidence for the accuracy of the employed 
numerical methods when this kink is visible in the $\pi^+\pi^-$ mass spectra 
calculated from a large number of terms. It is also interesting to consider 
the effects of the resummed Coulomb singularity by evaluating eq.\,(32) with 
the integrand $\mp(1-G)/\pi\alpha$, where $G=\eta/(e^\eta-1)$ is the Gamow 
function \cite{batgamow} with  $\eta=\mp 2\pi \alpha/\sqrt{s_1-4}$. In this 
treatment the dashed and dashed-dotted curves in Fig.\,12 result from the 
(singular part of the) Coulomb interaction between two pions with opposite 
and equal charges, respectively. In order to make the higher order 
electromagnetic effects better visible the fine-structure constant $\alpha $ 
has been increased by a factor 10 to $\alpha =1/13.7$. One observes that in the 
attractive case $(-)$ the kink in the dipion mass spectrum becomes more pronounced 
whereas in the repulsive case $(+)$ it gets apparently smoothened out. 

\begin{figure}[t!]
\begin{center}
\includegraphics[scale=0.5,clip]{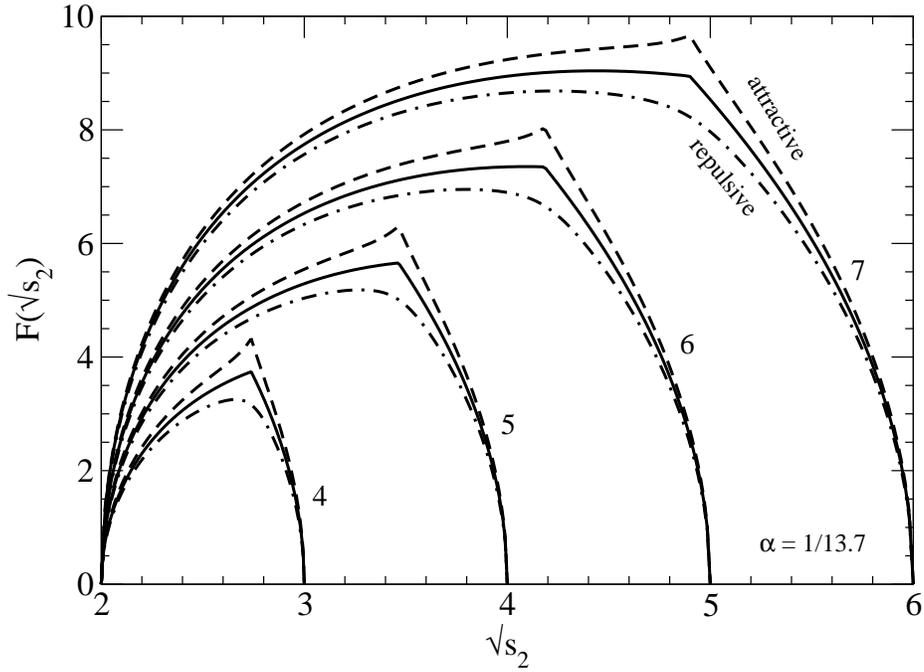}
\end{center}
\vspace{-.6cm}
\caption{Kink in the dipion mass spectrum at $s_2 = (s-1)/2$ caused by the 
Coulomb singularity in the final-state interaction.}
\end{figure}

\begin{figure}[t!]
\begin{center}
\includegraphics[scale=0.5,clip]{specminmin.eps}
\end{center}
\vspace{-.5cm}
\caption{$\pi^-\pi^-$ mass spectra of the process $\pi^-\gamma \to \pi^+ 
\pi^-\pi^-$ as a function of the $\pi^-\pi^-$ invariant mass $\mu \,m_\pi $ for 
values of $\sqrt{s}=4,5,6,7$.} 
\end{figure}

\begin{figure}[t!]
\begin{center}
\includegraphics[scale=1.0,clip]{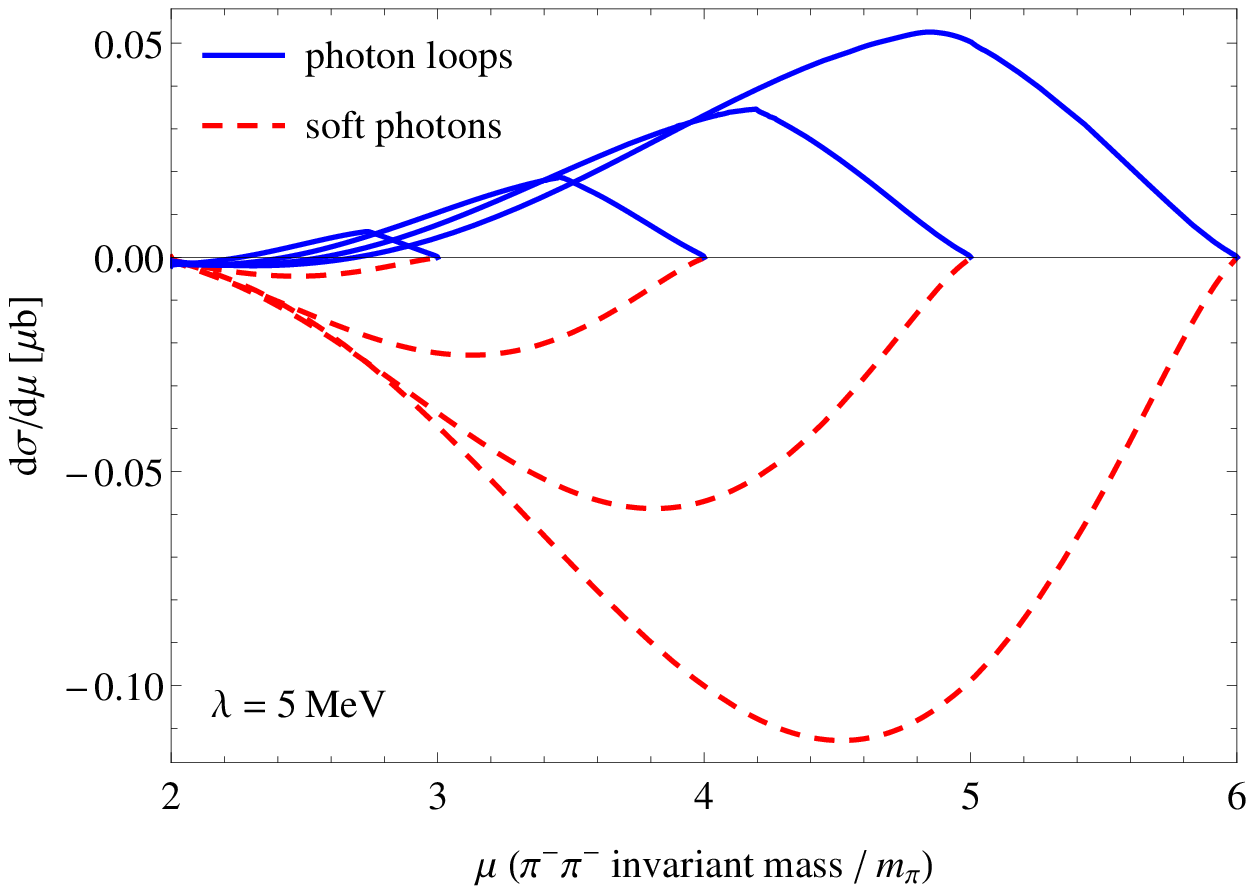}
\end{center}
\vspace{-.6cm}
\caption{Radiative corrections to the $\pi^-\pi^-$ mass spectrum of the process 
$\pi^-\gamma \to \pi^+ \pi^-\pi^-$ for $\sqrt{s}=4,5,6,7$.}
\end{figure}
\begin{figure}[h!]
\begin{center}
\includegraphics[scale=1.0,clip]{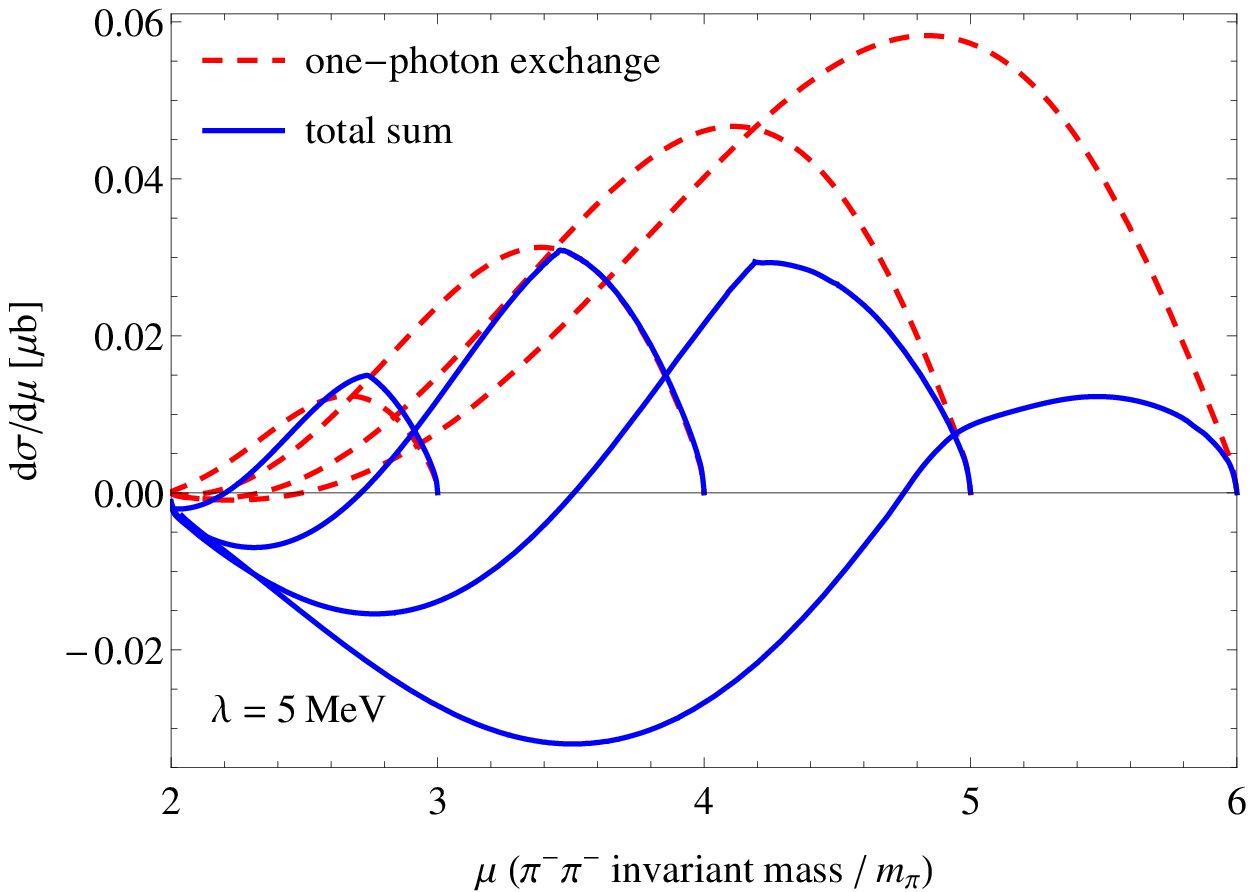}
\end{center}
\vspace{-.6cm}
\caption{Radiative corrections to the $\pi^-\pi^-$ mass spectrum of the process 
$\pi^-\gamma \to \pi^+ \pi^-\pi^-$ for $\sqrt{s}=4,5,6,7$.}
\end{figure}
We continue with the discussion of the $\pi^-\pi^-$ mass spectrum of the 
process $\pi^- \gamma \to \pi^+\pi^-\pi^-$. 
The $\pi^-\pi^-$ invariant mass is denoted by $\mu\,m_\pi $, where the 
(dimensionless) variable $\mu$ fulfills the relation $\mu^2 = s+3-s_1-s_2= 1-s+2 
\sqrt{s}(\omega_1+\omega_2)$. We introduce the sum $ \omega_+=\omega_1+\omega_2$ 
and half-difference $ \omega_-=(\omega_1-\omega_2)/2$. After this change of 
variables, $\omega_{1,2}=\omega_+/2\pm \omega_-$,  the differential cross section
$d\sigma/d\mu$ is obtained from the right 
hand side of eq.\,(5) by leaving out the integration over $\omega_+$ and 
multiplying with a factor $\mu /\sqrt{s}$. The limits for the remaining integration 
over $\omega_-$ are $\pm \sqrt{\mu^2-4} \sqrt{[s-(\mu+1)^2][s-(\mu-1)^2]}
/(4\mu \sqrt{s})$. For the purpose of comparison the $\pi^-\pi^-$ mass 
spectra calculated in tree approximation are reproduced in Fig.\,13. Note that 
these have a completely different shape than the $\pi^+\pi^-$ mass spectra 
displayed in Fig.\,9. The radiative corrections to the differential cross section 
$d\sigma/d\mu$ as they arise from soft-photon bremsstrahlung, virtual photon-loop 
and the one-photon exchange are shown in Figs.\,14,\,15, 
together with the total sum of these three contributions. The kinks at $\mu = 
\sqrt{(s-1)/2}$ caused by the Coulomb singularity in the $\pi^+\pi^-$ final-state 
interaction appear very prominently in Fig.\,15. The overlying bands produced by 
the variation of the low-energy constants  $\widehat k_1$ and $\widehat k_2$ have 
been omitted for reasons of a cleaner presentation. 

Altogether, the radiative corrections to the total cross section and dipion mass 
spectra of the reaction $\pi^-\gamma \to \pi^+ \pi^- \pi^-$ are of the order of a 
few percent, with the exception of the region $3<\sqrt{s}<4$ close to threshold. 
The electromagnetic corrections are indeed below the $5\%$ level as assumed in 
the analysis of the COMPASS data in ref.\,\cite{compass}. However, with the results 
of the present work the radiative corrections (as well as the leading isospin-breaking effect) 
can be consistently taken into account in the analysis of a future high statistics experiment.    
\section*{Appendix: Virtual radiative corrections to elastic 
{\boldmath$\pi^- \pi^-$} scattering}
In this appendix we study the virtual radiative corrections to elastic 
$\pi^- \pi^-$ scattering. Since $\pi^- \pi^- \to  \pi^- \pi^-$ is the strong 
interaction process underlying the charged pion-pair production $\pi^- \gamma 
\to \pi^+\pi^-\pi^-$, it is most instructive to investigate separately the 
radiative corrections for this simpler subprocess. An essential advantage is 
that these can be given in closed analytical form. Based on the chiral 
$\pi\pi$ contact-vertex at leading order, there are 10 associated photon-loop 
diagrams. For the four diagrams with external self-energy corrections the  
chiral tree amplitude $T_-^{(\rm tree)} =m_\pi^2(2-s)/f_\pi^2$ in the isospin limit gets 
multiplied with twice the (electromagnetic) wave function renormalization factor of 
the charged pion:
\begin{equation}T_-^{(\rm I)} = {2 \alpha \,m_\pi^2 \over \pi f_\pi^2} (2-s)
(\xi_{IR} -\xi_{UV})\,. \end{equation}
The two (equal) diagrams with a photon-loop in the $s$-channel (either in the initial or 
the final state) give together rise to a contribution to the $\pi^-\pi^-$ scattering 
amplitude, whose real part reads:
\begin{eqnarray} {\rm Re}\,T_-^{(\rm II)}&=&{\alpha\, m_\pi^2 \over \pi f_\pi^2} 
(2-s)\Bigg\{4\xi_{UV}-{9 \over 2} +{4-s+4 (s-2)\xi_{IR} \over \sqrt{s^2-4s}} 
\nonumber \\ &&\times \ln {\sqrt{s-4}+\sqrt{s} \over 2} +(2-s) 
-\hspace{-0.47cm}\int_4^\infty\! {dx \over x-s}\, { \ln(x-4) \over \sqrt{x^2-4x}} 
\Bigg\}\,. \end{eqnarray}
Here, we have used the (concise) spectral function representation of the scalar 
loop integral involving one photon and two pion propagators. 
In the same way one finds from the two photon-loop diagrams in the $t$-channel:
\begin{eqnarray} T_-^{(\rm III)}&=&{\alpha\, m_\pi^2 \over \pi f_\pi^2}\Bigg\{ 
\xi_{UV}\bigg( s+{3t \over 2}-5 \bigg) +{13-3s \over 2} -2t +\bigg[ 4s t +{3t^2 
\over 2} -8s\nonumber \\ &&  -10 t +16 +4(s-2)(2-t)\xi_{IR}\bigg] {1 \over 
\sqrt{t^2-4t}} \ln {\sqrt{4-t}+\sqrt{-t} \over 2} \nonumber \\ && +(s-2)(2-t) 
\int_4^\infty\! {dx \over x-t}\, { \ln(x-4) \over \sqrt{x^2-4x}} \Bigg\}\,, 
\end{eqnarray}
and the contribution from the photon-loops in the $u$-channel is immediately 
obtained via the substitution $t\to u$:
 \begin{eqnarray} T_-^{(\rm IV)}&=&{\alpha\, m_\pi^2 \over \pi f_\pi^2}\Bigg\{ 
\xi_{UV}\bigg( s+{3u \over 2}-5 \bigg) +{13-3s \over 2} -2u +\bigg[ 4s u +{3u^2 
\over 2} -8s\nonumber \\ &&  -10 u +16 +4(s-2)(2-u)\xi_{IR}\bigg] {1 \over 
\sqrt{u^2-4u}} \ln {\sqrt{4-u}+\sqrt{-u} \over 2} \nonumber \\ && +(s-2)(2-u) 
\int_4^\infty\! {dx \over x-u}\, { \ln(x-4) \over \sqrt{x^2-4x}} \Bigg\}\,. 
\end{eqnarray}

\begin{figure}[t!]
\begin{center}
\includegraphics[scale=0.5,clip]{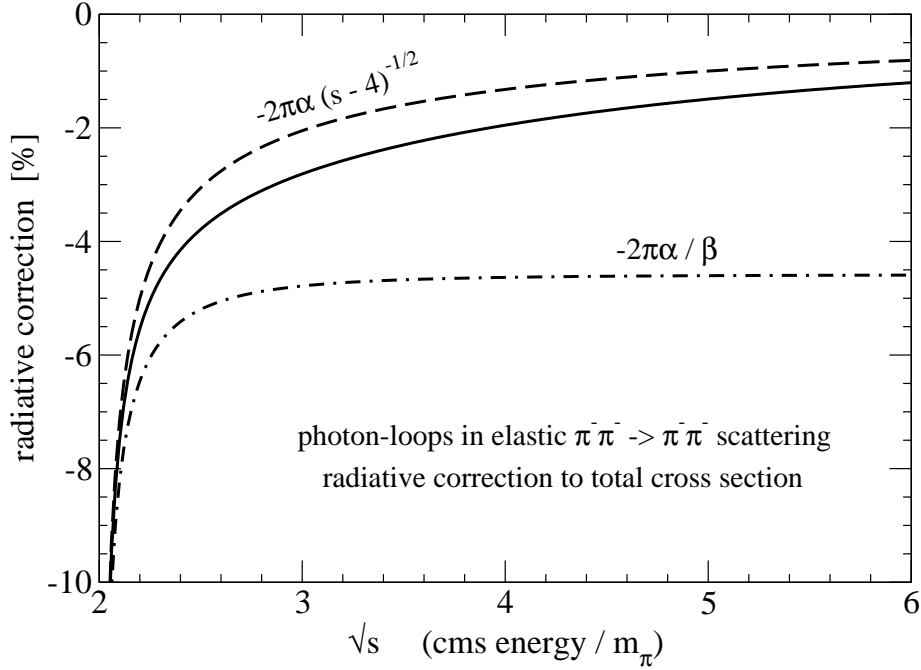}
\end{center}
\vspace{-.6cm}
\caption{Radiative corrections to the total cross section 
$\sigma_{\rm tot}(s)$ for elastic $\pi^-\pi^-$ scattering arising 
from virtual photon-loops.}
\end{figure}

Note that we have expressed here the elastic $\pi^-\pi^-$ scattering amplitude 
$T_-$ in terms of dimensionless Mandelstam variables $(s,t,u)$, which satisfy 
the constraint $s+t+u= 4$ and the inequalities $s>4$ and $t,u<0$ hold in the 
physical region \cite{cola}. The total ultraviolet divergence resulting from the 
sum of all 
10 photon-loop diagrams is: $T_-^{(\rm UV-div)}=-(\alpha\, m_\pi^2 /2\pi f_\pi^2)\, 
3s\,  \xi_{UV}$. This piece gets eliminated by the electromagnetic counterterms of 
ref.\,\cite{knecht}. The corresponding contribution to the $\pi^-\pi^-$ scattering 
amplitude reads:
\begin{equation}T_-^{(\rm ct)} = {\alpha \,m_\pi^2 \over 2\pi f_\pi^2} \Big[
( 3\xi_{UV}+\widehat k_1 )s + \widehat k_2  \Big]\,, \end{equation}
with $\widehat k_1$ and $\widehat k_2$ the same linear combinations of low-energy 
constants as written in eqs.\,(18,19). Let us note that the virtual photon-loops 
in charged pion-pion scattering have been calculated earlier by Knecht and Nehme 
\cite{nehme}. We find perfect agreement with the pertinent terms proportional 
to $e^2$ written in eqs.\,(12,13) of ref.\,\cite{nehme}. The infrared divergence is 
identified as $\xi_{IR} = \ln(m_\pi/m_\gamma)$ with $m_\gamma$ a regulator photon mass. 
The detailed comparison shows also that the term proportional to the low-energy constant 
$\widehat k_2$ in eq.\,(37) needs to be split up into two pieces and one part has 
been used in ref.\,\cite{nehme} to express the chiral tree amplitude $T_-^{(\rm tree)} 
=[m_\pi^2(4-s)-2m^2_{\pi^0}]/f_\pi^2$ in terms of the physical neutral pion mass.

Of particular interest is the threshold behavior 
of the last loop-function appearing in eq.\,(34). By expanding this principal-value 
integral around $s=4$ for $s>4$ one finds:
\begin{eqnarray} && -\hspace{-0.46cm}\int_4^\infty\! {dx \over x-s}\, { \ln(x-4) 
\over \sqrt{x^2-4x}} = {\pi^2 \over 2 \sqrt{s-4}} -1 -{\pi^2 \over 16}
\sqrt{s-4}+{\cal O}(s-4) \nonumber \\ && ={\pi^2 \over 4 v}(1-v^2) -1 
+{\cal O}(v^2)= {\pi^2 \over 4 \beta}(2-\beta^2) -1 
+{\cal O}(\beta^2) \,,  \end{eqnarray}
where the first term corresponds to the well-known Coulomb singularity 
proportional to the inverse pion velocity $v = \sqrt{1-4/s}$ in the 
center-of-mass frame or the inverse 
relative velocity $\beta = 2v/(1+v^2)= \sqrt{s(s-4)}/ (s-2)$. In 
non-relativistic quantum mechanics the same electromagnetic initial- or 
final-state 
interaction effect is described by the Gamow factor $\pm \pi \alpha/\beta $. 

The full line in Fig.\,16 shows the radiative corrections to the total cross 
section $\sigma_{\rm tot}(s)=(64\pi s\,m_\pi^2)^{-1} \int_{-1}^1\!dz\,|T_-|^2$ for 
elastic $\pi^-\pi^-$ scattering as they arise 
from the virtual photon-loops calculated in eqs.\,(34-36). We have discarded the 
$\xi_{UV}$ and $\xi_{IR}$ terms which get eliminated by the electromagnetic 
counterterm and the soft-photon bremsstrahlung. One observes negative corrections 
of the order of a several percent in the region $2<\sqrt{s}<6$. The behavior close to 
threshold is governed by the Coulomb singularity $-2\pi \alpha(s-4)^{-1/2}$, which 
is shown separately by the dashed line in Fig.\,16. The dashed-dotted line gives 
for comparison the (repulsive) Gamow factor $-2\pi \alpha/\beta$ with $\beta = 
\sqrt{s(s-4)}/ (s-2)$ the relative velocity of both pions. One is instructed that 
the result of the complete calculation in relativistic quantum field theory lies 
in between these two approximations. Interestingly, the pure 
Coulomb singularity $-2\pi \alpha(s-4)^{-1/2}$ provides the better approximation. 
Although the sign and $\sqrt{s}$-dependence are different, the 
magnitude of these virtual radiative corrections is comparable to the ones shown by 
the full line in Fig.\,7. Note that the one-photon exchange amplitude 
$T_-^{(1\gamma)}= 4\pi \alpha[(s-u)t^{-1}+(s-t)u^{-1}]$ for $\pi^-\pi^-\to\pi^-\pi^-$ has 
not been considered here since it would spoil the discussion in terms of the 
total cross section.

\end{document}